\documentclass[aps,pra,preprint,groupedaddress,showpacs]{revtex4-1}
\usepackage{feynmp}
\usepackage[pdftex]{graphicx}
\usepackage{braket}
\usepackage{amsbsy}
\usepackage{amsmath}
\usepackage{amssymb}
\usepackage{bm}
\usepackage{euscript}
\begin{document}

\title{Method for finding mechanism and activation energy of magnetic transitions, applied to skyrmion and antivortex annihilation}

\author{Pavel F. Bessarab$^{1,2}$, Valery M. Uzdin$^{2,3}$, and Hannes J\'onsson$^{4,5}$}
\affiliation{$^1$Dept. of Materials and Nanophysics, Electrum 229, Royal Institute of Technology (KTH), SE-16440 Kista, Sweden}
\affiliation{$^2$Dept. of Physics, St. Petersburg State University, St. Petersburg, 198504 Russia}
\affiliation{$^3$St. Petersburg National Research University of Information Technologies, Mechanics and Optics, St. Petersburg, 197101 Russia}
\affiliation{$^4$Faculty of Physical Sciences, VR-III, University of Iceland, Reykjav\'ik, Iceland}
\affiliation{$^5$Dept. of Applied Physics, Aalto University, FIN-00076 Espoo, Finland}

\date{\today}
\begin{abstract}
A method for finding minimum energy paths of transitions in magnetic systems is presented and used to 
determine the mechanism and estimate the activation energy of skyrmion and antivortex annihiliation in nano-systems. 
The path is optimized with respect to orientation of the magnetic vectors while 
their magnitudes are fixed or obtained from separate calculations. 
The curvature of the configuration space is taken into account by: 
(1) using geodesics to evaluate distances and displacements of the system during the optimization, and
(2) projecting the path tangent and the magnetic force on the tangent space of the manifold defined by all possible 
orientations of the magnetic vectors.
The method, named geodesic nudged elastic band (GNEB), and its implementation are illustrated with calculations 
of complex transitions involving annihilation and creation of skyrmion and antivortex states.
The lifetime of the latter was determined within harmonic transition state theory using a noncollinear extension of the 
Alexander-Anderson model.
\end{abstract}


\maketitle


\section{Introduction}

The assessment of the stability of magnetic states with respect to thermal fluctuations is an important problem 
in the theory of magnetism. The preparation of a magnetic system in a particular state can be destroyed by thermally activated transitions to other available states.
Thermal activation also needs to be taken into account when assessing the stability of a system with respect to external perturbations such as a magnetic field, contributing, for example, to the temperature dependence of hysteresis loops~\cite{brown_79}.

Thermal stability is a particularly important issue in the context of novel information storage devices. 
Magnetic skyrmions, for example, are localized, noncollinear spin configurations demonstrating particle-like behavior
and are promising candidates for future data storage due to their small size and topological stability.
They have been identified both in bulk magnetic materials~\cite{muhlbauer_09} and in 
thin magnetic films~\cite{heinze_11}.  The stability of skyrmions against thermal fluctuations is
an essential prerequisite for their use in memory devices. Originally, skyrmions were introduced in the context of elementary particles as configurations of continuous fields with topological charge. In a continuum limit, topological protection makes them stable against arbitrarily large fluctuations.
However, in real systems with magnetic moments localized on atoms, the topological protection is not strict. 
Instead, skyrmion states are separated from the topologically simple (e.g. ferromagnetic) states by finite activation barriers 
which define their stability and lifetime. 
Vortices and antivortices are other noncollinear spin configurations with potential applications in magnetic information technology and spintronics.
A methodology for accurately determining the mechanism and rate of annihilation of such noncollinear magnetic 
states is needed, for example, in the exploitation of these kinds of states in devices. 

Thermally activated magnetic transitions are typically rare events on the time scale of oscillations 
of the magnetic moments, 
making direct simulations of spin dynamics an impractical way to calculate transition rates.
This separation of time scales, however, makes it possible to apply statistical approaches such as transition state theory 
(TST)~\cite{TST} or Kramers theory~\cite{kramers_40}. 
Within the harmonic approximation to TST (HTST)~\cite{HTST}
and within Kramers theory, the activation energy of a transition, 
the primary quantity determining thermal stability, 
is given by the energy difference between the local minimum of the energy surface corresponding to the initial state and
the highest first order saddle point (SP) located on a path connecting the initial and final state minima. 
In adaptions of these rate theories to magnetic systems~\cite{brown_79,braun_94,visscher_12,fiedler_12,bessarab_12}, the magnitude of the magnetic vectors is either assumed to be constant as orientation changes, 
or it is treated as a fast variable obtained from self-consistency calculations for fixed values of the ÒslowÓ variables that 
specify the orientation~\cite{bessarab_14}.  The energy surface of a system of $N$ magnetic 
moments is then a function of $2N$ degrees of freedom defining the orientation of the magnetic moments.
The description of thermal stability of the magnetic states essentially becomes a problem of identifying the SPs on the energy surface. 
For a magnetic system with several degrees of freedom, locating the relevant SPs is the most challenging part of the calculation.  
The difficulty arises from the need to minimize the energy with respect to all but
one degree of freedom for which a maximization should be carried out.  
The problem is that it is not known
{\it a priori} which degree of freedom should be treated differently. The special degree of freedom for which a maximization
should be carried out is often referred to as the {\it reaction coordinate}.

An approach which is commonly used in studies of atomic rearrangements but is, however, not well justified, relies on 
some reaction coordinate chosen before the calculation is carried out. 
For a sequence of values of this coordinate, the energy is minimized with respect to all 
orthogonal degrees of freedom. This method goes by many names and is often reinvented. In calculations of atomic
systems it has, for example, been referred to as the `drag method' because the system is dragged stepwise along the 
assumed reaction coordinate while it relaxes in other degrees of freedom.  
The maximum energy obtained in this way is then taken to be the SP energy. 
This approach has been applied to studies of magnetization reversal in magnetic systems
where, for example, a component of the average magnetization vector of the system has been chosen as a reaction coordinate~\cite{paz_08}.  
The method works well if the choice of reaction coordinate happens to be a good one. It can, however, give 
inaccurate results if the assumed reaction coordinate turns out to be a poor choice. The calculated path traced out by 
this procedure may end up being discontinuous and the 
maximum energy obtained is then not an accurate estimate of the SP energy~\cite{NEBleri}.
In some cases, it is even hard to come up with a guess of a reaction coordinate.  One example of such a transition in a magnetic system
is the reversal of a vortex core~\cite{waeyenberge_06} which involves formation of a vortex-antivortex pair, 
a complex mechanism for which simple choices of a reaction coordinate would likely fail. 

A definitive identification of the relevant SP for a transition between given initial and final states involves finding the 
minimum energy path (MEP) between the corresponding local minima on the energy surface. 
Following an MEP means advancing each degree of freedom of the system in such a way that the energy is minimal with respect to all degrees
of freedom perpendicular to the path. An MEP is a path of highest statistical weight connecting the initial and final states and, thereby, represents a particular mechanism of the transition. 
There can be more than one MEP between given initial and
final states, each path corresponding to a specific mechanism and transition rate
that can be estimated using the HTST or Kramers approaches.
The tangent to the MEP provides a good choice of a reaction coordinate and the displacement along the MEP
represents the advancement of the system during the transition from the initial state to the final state 
following a specific mechanism. 
Any maximum on an MEP is a SP on the energy surface but it is the highest maximum that
gives an estimate of the activation energy for the corresponding transition mechanism. 

The first approach we are aware of for finding MEPs of magnetic transitions involves direct minimization of the 
thermodynamic action of the system~\cite{berkov_98}. 
The minimization of the action, however, can converge to any path that lies along gradient lines of the energy surface
and connects the two minima. 
As a result, an MEPs may not be found but rather a path passing through second- or higher-order SPs or even a local maximimum on the energy surface. 
Such `false' paths do not provide an estimate of the thermal activation energy. 
It is expected that the number of such irrelevant paths increases as the number of degrees of freedom in the system 
increases, making it difficult to apply the method to large and complex systems.

The nudged elastic band (NEB) method~\cite{mills_95,NEBleri} 
and variations thereof~\cite{henkelman_00,e_02,trygubenko_04,zarkevich_15}, 
are commonly used to find MEPs of transitions involving atomic rearrangements.
The method involves specifying a discrete representation of some initial path 
between the two minima and then using an iterative algorithm to bring the discretization points to the nearest MEP.
A guess of the reaction coordinate is not needed. Rather, it is determined from the iterative optimization of the path.    
The discretization points represent values of the full set of variables of the system and are referred to as `images' of the system.
The images provide a discrete approximation to a continuous path which should be displaced during the optimization only in directions perpendicular to the path~\cite{NEBleri}.
In order to enforce that, a tangent to the path needs to be estimated at each image 
and the magnetic force component in the direction of the tangent removed. 
In a practical calculation, the number of images is finite and it is important to control the distribution of the images along the 
path to get adequate resolution while keeping the number of images to a minimum and, 
thereby, minimizing the computational effort.
Good resolution is especially important near the SPs and in regions where the path has large curvature. 
The distribution of images along the path can be controlled by adding spring forces between adjacent images.
It is important to include only the component of the spring force pointing in the direction of the path tangent so as not 
to interfere with the displacement of the path towards the MEP during the optimization.  
This projection of the true force and the spring force using the tangent to the path is referred to as `nudging'. 
Without control over the distribution of images, the density of images will drop in regions near saddle points  
where good resolution is particularly important.  
An alternative way of distributing the images is to estimate the length of 
the path and redistribute the images at each iteration using an interpolation between current locations of the 
images~\cite{e_02}.
The NEB method has turned out to be a powerful tool for determining transition mechanisms of atomic rearrangements, 
such as chemical reactions and diffusion events. 
Complex and counterintuitive mechanisms have, for example, been found for catalytic reactions and 
diffusion events controlling the morphology of growing crystal surfaces~\cite{jonsson_11,villarba_94}.

The NEB method and variations of it have also been applied to magnetic systems
both on the atomic scale and on a mesoscopic scale described by micromagnetics~\cite{visscher_12,fiedler_12,bessarab_12,bessarab_14,dittrich_02,dittrich_03,thiaville_03,e_03,dittrich_04,dittrich_05,suess_06,goll_07,suess_07,berkov_07,krone_10,tudosa_12,bessarab_13}.
In many cases, these have been rather simple systems with 
the magnetic vectors rotating mainly in a plane 
(with some exceptions, for example~\cite{thiaville_03,dittrich_05,krone_10}). 
Several choices of coordinate systems are possible in such calculations. In some cases, spherical polar coordinates 
were used and the length of the magnetic vectors kept constant during the transitions~\cite{dittrich_02}. 
The distance between images in the NEB path was calculated by treating the spherical coordinates as if they are linear.
This is a rough approximation which leads to reduced
control of the distribution of the images along the path, especially when some images are in the vicinity of the poles. 
The evaluation of the spring force can also be problematic 
and the springs have in some cases been dropped~\cite{dittrich_02}. 
But, this is not a recommended practice and serves little purpose if the curvature of the 
configuration space is properly taken into account
since the evaluation of the spring force adds only an insignificant computational effort. 
In other applications, Cartesian coordinates have been used to describe the magnetic vectors~\cite{suess_07}. 
This makes the evaluation of the distance between images easier, 
but the contstraint on the magnitude of the magnetic vectors becomes more complicated. 
It can be enforced by projecting the NEB force on the plane perpendicular to the magnetic moments. 
It has been noted, however, that this approach can suffer from 
convergence problems where, for example, 
the choice of the value of the spring constant becomes critical~\cite{suess_07,berkov_07}, 
unlike NEB calculations of atomic rearrangements where the spring constant can be chosen 
over a range spanning many orders of magnitude~\cite{NEBleri,mills_95}.  
We show here that this apparent sensitivity to the value of the spring constant in calculations of magnetic transitions
is not present if the tangent to the path is evaluated properly, as described below.


In this article, an extension of the NEB method to magnetic systems, the geodesic NEB (GNEB), is presented and applied to complex transitions involving 3-dimensional rotations of the magnetic vectors.  
The modifications of the NEB needed for magnetic systems arise from the fact that the configuration space of a magnetic system is a curved manifold due to the constraints on the length of the magnetic vectors, which is either assumed to be 
constant or obtained from a separate, self-consistency calculation. 
As a result, several aspects of the NEB method have been modified, such as 
(1) the use of geodesics to evaluate the distance between adjacent images and to estimate displacements of the images 
during the optimization, and 
(2) projection of the path tangent as well as the force on the tangent space of the manifold of magnetic configurations defined by all possible orientations of the magnetic vectors.
While various coordinate systems can be used with the GNEB method, 
we have chosen here to use spherical polar coordinates. A special treatment of the coordinates is then needed close to the poles.
Also, while various minimization methods can be used in conjunction with the GNEB method, 
we present here an algorithm based on an equation of motion, analogous to a robust minimization algorithm that has been used extensively with the NEB method~\cite{NEBleri}. 
The GNEB method can be applied to complex magnetic transitions, as is demonstrated here in calculations of skyrmion and antivortex annihilation.
The value chosen for the spring constant turns out not to be critical, a wide range of values can be used, 
analogous to the NEB method for atomic rearrangements.


The article is organized as follows. 
A review of the NEB method for atomic transitions is given in Sec. II. The GNEB method is then presented in Sec. III.
Results of applications of the GNEB method are given in Sec. IV, first using a single spin test system 
to illustrate the importance of properly projecting the path tangent as well as problems that can arise if springs are not included.
Then, the application to skyrmion and antivortex annihilation is described. 
A discussion and summary is presented in Sec. V.  
Seven Appendices give additional technical information: 
(A) equations for evaluating the path tangent;
(B) equations for evaluating the geodesic length;
(C) detailed description of the Hamiltonian used in the single spin test problems;
(D) an optimization algorithm based on velocity projection;
(E) a method for interpolating the energy using a discrete representation of a path;   
(F) a method for verifying that a first order saddle point has been identified; and
(G) a brief, step-by-step summary of a GNEB calculation. 


\section{NEB method}
\label{sec:neb}

The basic equations of the NEB are reviewed here in order to make the subsequent section on the 
extensions made in the GNEB method clearer. 
The NEB method was developed to find MEPs of transitions where coordinates of atoms change in, for example, chemical 
reactions, diffusion events or changes in molecular configuration. 
The location of each particle is given by a 3-dimensional vector, $\vec r$, 
and a system of $N$ interacting particles is desribed by a $3N$-dimensional vector in Euclidean space
$\bm{R} = \left(\vec {r}_1, \vec {r}_2, \dots \vec {r}_N\right)$. 
The fact that the configurational space is Euclidean is used explicitly in the calculation of the distance between images, 
to estimate the local tangent to the path and to project the forces.

Consider a chain of $Q$ images, $\left[\bm{R}^1,\bm{R}^2,\ldots,\bm{R}^Q\right]$, where the endpoints are fixed 
and given by the initial and final configurations of the transition, but $Q-2$ intermediate images 
$\bm{R}^\nu=\left(\vec{r}_1^\nu,\vec{r}_2^\nu,\ldots,\vec{r}_N^\nu\right)$, where $\nu=2,\ldots,Q-1$, 
give a discrete representation of a path. The position of the intermediate images
needs to be adjusted in order to converge on the MEP. This is accomplished by displacing the images
along the force acting on them so as to zero the force. 
The key idea of the NEB method is the force projection -- the nudging. 
It is based on the observation that only transverse displacements of the path should be included. 
A displacement of the images along the path is simply a discrete analog of a reparametrization of 
a continuous path~\cite{NEBleri} and does not affect the location of the path in configuration space. 
Any convenient distribution of the images along the path can be used as long as the resolution is high enough.
As for any numerical method involving discretization, the number of images needs to be large enough to reach convergence. 
Some reagions along the path may require higher resolution than others.
One can, for example, choose to increase the density of images in regions where the energy is high
or the curvature is large~\cite{henkelman_00}. 

The true force is the negative of the energy gradient, $\bm{F} =-\nabla E$.
The transverse displacements are generated by the component of the true force perpendicular to the path. 
Therefore, only this component should be included in the force when the images are displaced. 
Otherwise, the true force would move the images along the path 
towards regions of low energy and the information about region(s) near saddle point(s), the most important part(s) of the path, would be lost. 
Given an estimate of the unit tangent to the path at each image, $\hat{\bm{\tau}}^\nu$, the perpendicular component of the 
energy gradient is obtained by subtracting the parallel component:
\begin{equation}
\label{eq:perp}
\nabla E\left(\bm{R}^\nu\right)|_\perp = 
\nabla E\left(\bm{R}^\nu\right)-\left(\nabla E\left(\bm{R}^\nu\right)\cdot\hat{\bm{\tau}}^\nu \right)\hat{\bm{\tau}}^\nu .
\end{equation}

When the distribution of the images along the path is controlled by including springs between adjacent images,
the spring force has to be projected on the tangent to the path so as not to affect the transverse displacements of the images. 
The parallel component of the spring force can be evaluated as
\begin{equation}
\bm{F}^\nu_s|_\parallel = \kappa\bigl[\left| \bm{R}^{\nu+1}-\bm{R}^{\nu}\right|-\left|\bm{R}^{\nu}-\bm{R}^{\nu-1}\right|\bigr]\hat{\bm{\tau}}^\nu,
\end{equation}
where $\kappa$ is a spring constant and $| \bm{R}^{\nu+1}-\bm{R}^{\nu}|$ denotes the Euclidean distance 
between adjacent images $\nu+1$ and $\nu$.

The total NEB force is then evaluated as:
\begin{equation}
\label{eq:forceNEB}
\bm{F}^\nu = -\nabla E\left(\bm{R}^\nu\right)|_\perp+\bm{F}^\nu_s|_\parallel .
\end{equation}
%


In ordert to form the force projections, it is essential to have a good estimate of the 
tangent to the path at each image, $\hat{\bm{\tau}}^\nu$. 
The simplest approach is to use the line segment connecting the previous and later image in the path, $\nu+1$ and  $\nu-1$.  This, however,
has been found to lead to instabilities in some cases, slowing down or even preventing convergence~\cite{NEBleri}.  
A better way is to use the neighboring image that has higher energy in either a forward or backward finite difference scheme, 
and linearly switching between the two schemes when the energy of image $\nu$ is not in between the energy 
of its two neighbors~\cite{newtang}.

Some initial path is needed to start an NEB calculation. This can be done in many different ways, but the most commonly 
used method is to generate a linear interpolation between the endpoints,
$\bm{R}_i^\nu = \bm{R}_1 + (\nu-1) (\bm{R}_Q -\bm{R}_1)/(Q-1)$. 
A better starting configuration can be obtained by carrying out a procedure based on an 
interpolation of pairwise distances between atoms, 
the image dependent pair potential (IDPP) method~\cite{sorensen14}, since it prevents atoms from coming too close 
together and generally gives a path that is closer to an MEP.

When two or more MEPs connect the same initial and final configurations, 
the optimization procedure will most likely lead to convergence to the MEP closest to the initial path. 
In order to find the optimal MEP in such a situation, some sampling of the various MEPs needs to be carried out. 
Typically, an NEB calculations involves on the order of ten images.  When complex paths involving intermediate minima 
are calculated, it is advisable to divide the path at an intermediate minimum and carry out separate NEB calculation of 
each segment of the MEP. 

When the energy along a path is inspected, it is important to use information contained in the force along the path 
in the interpolation scheme as well as the energy at each image.  
This makes it easier to identify, for example, intermediate minima that might be present. 
An interpolation procedure that has proved to be useful in calculations of atomic rearrangements is presented in reference \cite{newtang}.

Typically, the most important result sought from an NEB calculation is the highest energy along the MEP since this gives
an estimate of the activation energy for the transition. The saddle point will, however, most likely lie in between NEB 
images and the estimate of the maximum energy then be subject to errors in the interpolation between images. 
In order to determine the maximum energy accurately, the highest energy image can be treated separately during the 
iterative NEB optimization and made to move uphill in energy along the path.
The force on this climbing image (CI) is calculated by zeroing the spring force acting on it and 
inverting the parallel component of the true force~\cite{henkelman_00}
\begin{equation}
\label{eq:forceCI}
\bm{F}^{\nu_{CI}} = -\nabla E\left(\bm{R}^{\nu_{CI}}\right)+2\bigl(\nabla E\left(\bm{R}^{\nu_{CI}}\right)\cdot\hat{\bm{\tau}}^{\nu_{CI}}\bigr)\hat{\bm{\tau}}^{\nu_{CI}} .
\end{equation}
In this way, the climbing image is made to move uphill in energy along the path but downhill in energy perpendicular to the path. 
The adjacent images effectively define the reaction coordinate and, thereby, the appropriate climbing direction to reach a first 
order saddle point. 
After the CI-NEB calculation has converged, the position of the climbing image coincides with the highest SP 
along the MEP and gives an accurate value of the saddle point energy.


\section{Geodesic NEB method}
\label{sec:gneb}

The state of a magnetic system consisting of $N$ magnetic moments is in principle specified by $3N$ parameters 
-- the components of the magnetic vectors in 3-dimensional space.
However, changes in the magnitude of the magnetic momentum associated with each magnetic atom 
are in most materials much faster than changes in orientation~\cite{antropov_96}. Therefore, the local magnetization and, 
thereby, the energy of the system can usually be described as
a function of the orientation of the magnetic vectors only. This is analogous to the Born-Oppenheimer approximation in atomic systems where the electronic degrees of freedom are assumed to be fast compared with the slowly varying 
positions of the nuclei. The total energy of the system 
can then be expressed as a function of only the slow degrees of freedom.
As a result, the configurational space of a system of $N$ magnetic moments contains $N$ constraints on the magnitude of the vectors. 
The NEB method can still be applied in such cases, but the constraints would be violated in each iteration and need to be restored in some way.
Such a 'constrained' NEB method can suffer from slow convergence similar to what has been reported in analogous constrained optimization problems (see, for example, ref.~\cite{abrudan_08} and references therein). An alternative is to formulate the 
calculation as an unconstrained optimizattion in properly chosen configuration space. 
For magnetic systems, instead of $3N$-dimensional Euclidean space with $N$ constraints, one can choose a 
$2N$-dimensional Riemannian manifold, $\mathcal{R}$, 
corresponding to the direct product of $N$ 2-dimensional spheres: 
\begin{equation}
\label{eq:conf_space}
\mathcal{R} = \prod_{i=1}^N S_i^2,
\end{equation}
where $S_i^2$ is a $2$-dimensional unit sphere associated with the $i$-th magnetic momentum vector. 
The basic equations used in the NEB method need to be revised for this case because the 
Riemannian manifold is a curved space. 
The GNEB method presented here is an unconstrained optimization method within the $\mathcal{R}$-manifold.

Similar to the NEB method, the GNEB method involves a chain of images of the system giving a discrete representation of a path
between the local energy minima corresponding to the initial and final states. Adjacent images are connected with 
springs in order to ensure continuity of the path, as in the NEB method. 
At each image, a local tangent to the path needs to be estimated 
and the force guiding the images towards the nearest MEP is defined 
as the sum of the transverse component of negative energy gradient,
i.e. the true force, plus the component of the spring force along the tangent.   
The position of intermediate images is then adjusted so as to zero the GNEB force. 

A particularly important, but non-trivial aspect of the GNEB method is the projection of the tangent on the tangent space of the Riemannian manifold.
Without it, the true force and the spring force
interfere with each other leading to uncontrolled behavior of the path, as explained below.
The tangent space to the $\mathcal{R}$-manifold can be illustrated as follows. 
Consider a single spin system where the state is described by a unit 
vector $\vec{m}$ corresponding to a point on a 2{\it D} unit sphere, $S^2$. 
The tangent space $\mathcal{T}_{\vec{m}}S^2$ to $S^2$ at a point $\vec{m}\in S^2$ is defined as 
the span of all vectors $\vec{X}$ perpendicular to $\vec{m}$:
\begin{equation}
\label{eq:tangentS}
\mathcal{T}_{\vec{m}}S^2 = \bigl\{\vec{X}\colon\vec{X}\cdot\vec{m}=0\bigr\}.
\end{equation}
For a system of $N$ spins, the tangent space $\mathcal{T}_{\bm{m}}\mathcal{R}$ to the $\mathcal{R}$-manifold at a point 
$\bm{m}=\left(\vec{m}_1,\vec{m}_2,\ldots,\vec{m}_N\right)$ 
is a direct product of tangent spaces to the unit spheres $S^2_i$ 
associated with each unit vector $\vec{m}_i$ specifying the orientation of the $i$-th magnetic moment:
\begin{equation}
\label{eq:tangentC}
\mathcal{T}_{\bm{m}}\mathcal{R} = \prod_{i=1}^N \mathcal{T}_{\vec{m}_i}S^2_i.
\end{equation}

A projection $ \mathcal{P}_\mathcal{T} \bm{A}$ of any 3$N$-dimensional vector 
$\bm{A} = \bigl(\vec{A}_1,\vec{A}_2,\ldots,\vec{A}_N\bigr)$ on 
the tangent space $\mathcal{T}_{\bm{m}}\mathcal{R}$ can be obtained by subtracting out the component parallel to 
$\vec{m}_i$ from each 3{\it D}-vector $\vec{A}_i$:
\begin{equation}
\label{eq:projS}
\vec{A}_{i,\mathcal{T}} = \vec{A}_i-\bigl(\vec{A}_i\cdot\vec{m}_i\bigr)\vec{m}_i.
\end{equation}
$ \mathcal{P}_\mathcal{T} \bm{A}$ is then a composition of $\vec{A}_{1,\mathcal{T}},\ldots,\vec{A}_{N,\mathcal{T}}$:
\begin{equation}
\label{eq:projC}
 \mathcal{P}_\mathcal{T} \bm{A} = \bigl(\vec{A}_{1,\mathcal{T}},\vec{A}_{2,\mathcal{T}},\ldots,\vec{A}_{N,\mathcal{T}}\bigr).
\end{equation}

With the definitions given in Eqs.~(\ref{eq:tangentS})-(\ref{eq:projC}), the algorithm involved in the GNEB method can now be described. 
A chain of $Q$ images is constructed, $\left[\bm{M}^1,\bm{M}^2,\ldots,\bm{M}^Q\right]$, 
where the endpoints are fixed and given by the 
local minima corresponding to the initial and final states, while the location of the $Q-2$ intermediate images 
$\bm{M}^\nu=\left(\vec{m}_1^\nu,\vec{m}_2^\nu,\ldots,\vec{m}_N^\nu\right)$, $\nu=2,\ldots,Q-1$, 
is adjusted so as to zero the GNEB force acting on them. 
The basic force is defined in a way analogous to the NEB force, as in Eqn.(\ref{eq:forceNEB}), with $E(\bm{R}^\nu)$ replaced by $E(\bm{M}^\nu)$.
But, now the tangent used to project the true force and the spring force has to be projected on the 
tangent space of the $\mathcal{R}$-manifold.
First, the tangent is estimated in a way analogous to the NEB method, see Appendix~\ref{app:path_tang}, and then it is projected on the tangent space
\begin{equation}
\label{eq:TanProj}
\hat{\bm{\tau}}^\nu_\mathcal{T} = \frac{ \mathcal{P}_\mathcal{T} \hat{\bm{\tau}}^\nu}{|  \mathcal{P}_\mathcal{T} \hat{\bm{\tau}}^\nu  |} .
\end{equation}
The perpendicular component of the energy gradient can then be obtained by subtracting out the parallel component
\begin{equation}
\label{eq:Gperp}
\nabla E\left(\bm{M}^\nu\right)|_\perp = \nabla E\left(\bm{M}^\nu\right)-\left(\nabla E\left(\bm{M}^\nu\right)\cdot\hat{\bm{\tau}}^\nu_\mathcal{T}\right)\hat{\bm{\tau}}^\nu_\mathcal{T} .
\end{equation}
The parallel component of the spring force is evaluated using geodesic distances
\begin{equation}
\label{eq:GFsparallel}
\bm{F}^\nu_s|_\parallel = \kappa\left[L\left(\bm{M}^{\nu+1},\bm{M}^{\nu}\right)-L\left(\bm{M}^{\nu},\bm{M}^{\nu-1}\right)\right]\hat{\bm{\tau}}^\nu_\mathcal{T}.
\end{equation}
Here, $L\left(\bm{M}^{\nu+1},\bm{M}^{\nu}\right)$ and $L\left(\bm{M}^{\nu},\bm{M}^{\nu-1}\right)$ are geodesic distances between images $\nu+1$, $\nu$ and $\nu$, $\nu-1$, respectively. 
Since the $\mathcal{R}$-manifold is equal to a direct product of $S^2_i$ (see Eq.~(\ref{eq:conf_space})), 
the distance between points $\nu$ and $\mu$ is given by:
\begin{equation}
\label{eq:geodesicdist}
L\left(\bm{M}^{\nu},\bm{M}^{\mu}\right) = \sqrt{l^{\nu,\mu}_1+l^{\nu,\mu}_2+\ldots+l^{\nu,\mu}_N}.
\end{equation}
Here $l^{\nu,\mu}_i$ are great-circle distances between points $\nu$ and $\mu$ on the $i$-th unit sphere. 
$l^{\nu,\mu}_i$ can be computed using, for example, spherical law of cosines, haversine or Vincenty's formulae~\cite{korn_00, sinnott_84, vincenty_75} (see Appendix~\ref{app:geo_dist}).
Here, the same spring constant, $\kappa$, is used in the applications of the method, 
but a non-uniform distribution of the images along the path can be 
obtained by choosing different values of the spring constant for each geodesic 
segment, using stiffer springs in regions where higher resolution of the path is desired.

The force $\bm{F}^\nu$ obtained from Eq.~(\ref{eq:forceNEB}) by inserting $\nabla E\left(\bm{M}^\nu\right)|_\perp $ 
from Eq.~(\ref{eq:Gperp}) and
$\bm{F}^\nu_s|_\parallel$ from Eq.~(\ref{eq:GFsparallel})
is in general not in the tangent space of $\mathcal{R}$
even though the tangent, $\hat{\bm{\tau}}^\nu_\mathcal{T} \in \mathcal{T}_{\bm{m}^\nu}\mathcal{R}$, is. 
The force that should be applied to the images in the GNEB method, therefore needs to be 
projected on the tangent space to satisfy the magnetic constraints:
\begin{equation}
\label{eq:forceGNEB}
\bm{F}^\nu_{GNEB} = \mathcal{P}_\mathcal{T} ( - \nabla E\left(\bm{M}^\nu\right)|_\perp + \bm{F}^\nu_s|_\parallel)
\end{equation}
where the projection, $\mathcal{P}_\mathcal{T} $, is calculated using Eqs.~(\ref{eq:projS}) and~(\ref{eq:projC}). 
The GNEB optimization procedure iteratively zeros the force $\bm{F}^\nu_{GNEB}$ to bring the images to the nearest MEP.

\begin{figure}[h!]
\centering
\includegraphics[width=0.35\columnwidth]{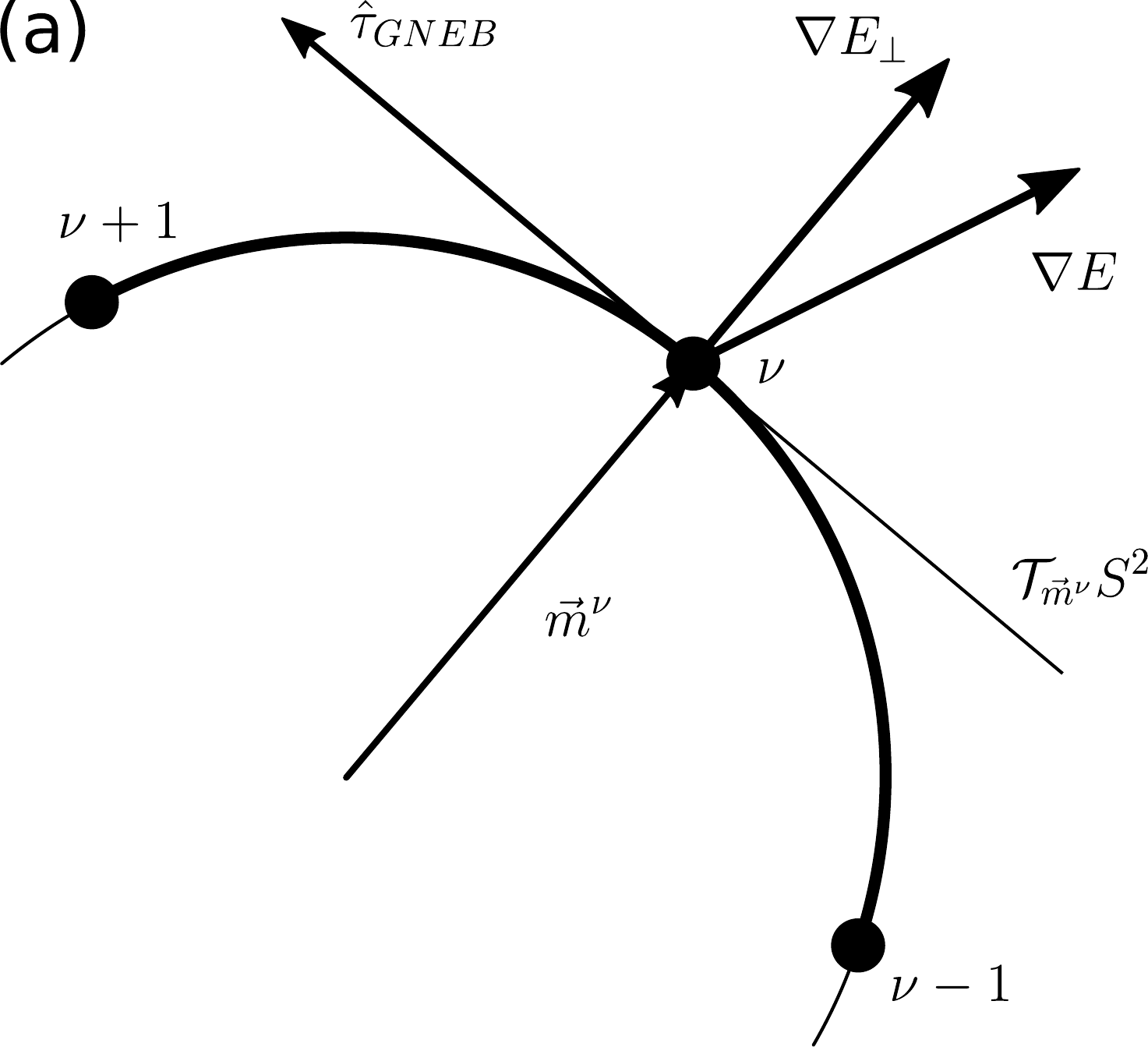}
\hskip 1 true cm
\includegraphics[width=0.35\columnwidth]{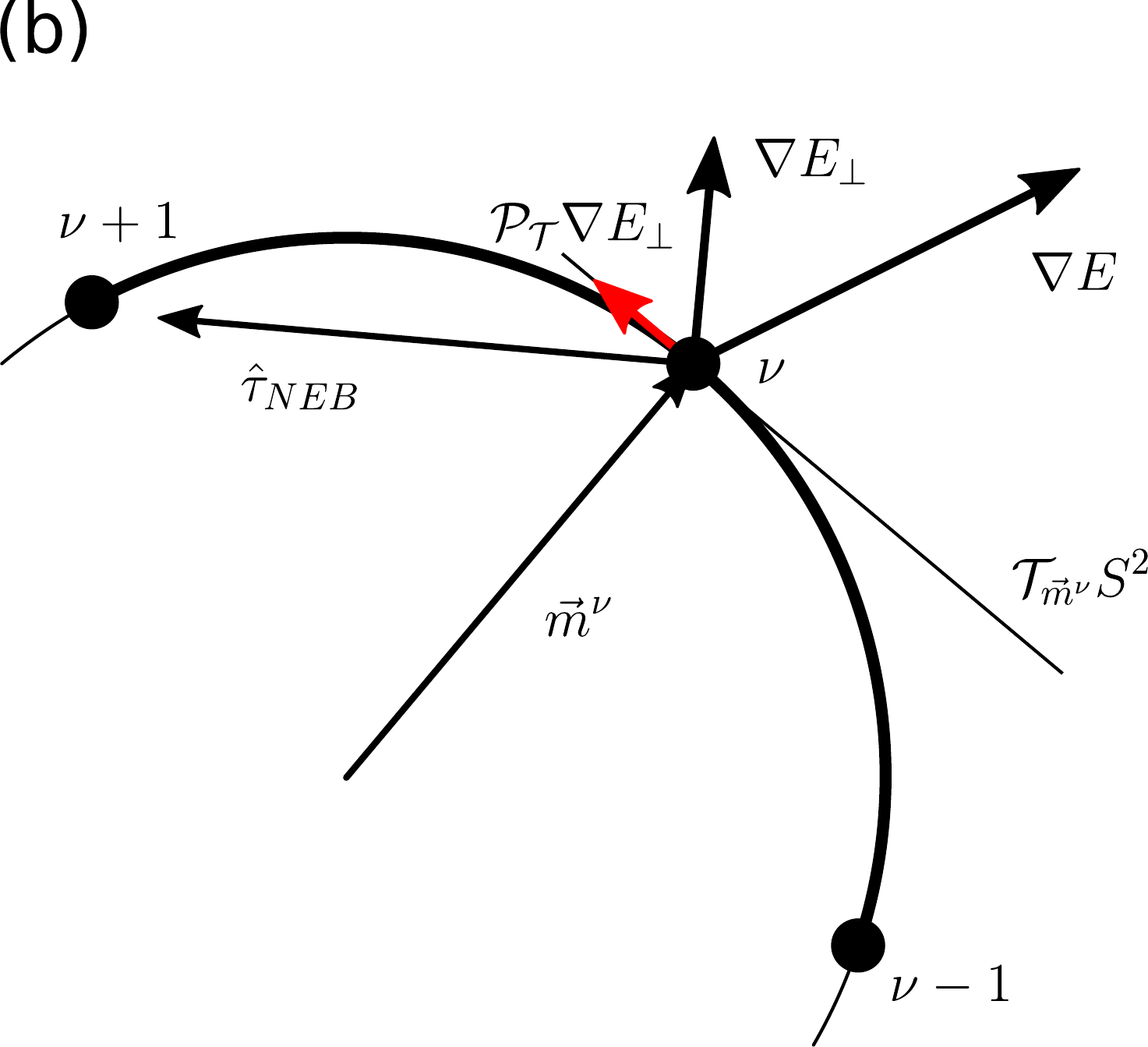}
\caption{
Illustration of the force projection used in the GNEB method and the error arising if the tangent is not projected
on the tangent space. Three equidistant images, $\nu-1$, $\nu$ and $\nu+1$, marked with filled circles, are located along 
a geodesic path lying in the plane of the figure. The NEB tangent to the path at image $\nu$, $\hat{\bm{\tau}}_{NEB}$, 
is defined as a unit vector pointing to a higher energy image $\nu+1$. 
(a) The tangent projected on the tangent space, $\hat{\bm{\tau}}_{GNEB} = \hat{\bm{\tau}}_\mathcal{T}$, is used to project 
out the parallel component of the gradient, $\nabla E$, 
to obtain the transverse component, $\nabla E_\perp$, which does not move
the image along the geodesic path. The distribution of images along the path is, thereby, not affected.
(b) A problem occurs if the tangent is not projected on the tangent space and $\hat{\bm{\tau}}_{NEB}$ is used to estimate 
the transverse component of $\nabla E$. 
A non-zero component of the force acting on the image remains in the tangent space, as shown with a
red arrow. The true force then interferes with the distribution of images along the path.
}
\label{fig1}
\end{figure}

The key difference between the tangent used in the GNEB method, $\hat{\bm{\tau}}^\nu_\mathcal{T}$, and the tangent used in the NEB method, $\hat{\bm{\tau}}^\nu$, 
is that the former lies in the tangent space of the $\mathcal{R}$-manifold of image $\nu$, see Fig. 1a.
This is needed for the true force to be decoupled from the spring force. 
If the NEB tangent is used for magnetic transitions, without projecting on the tangent space, 
interference occurs between the true force and the 
spring force and, thus, uncontrolled behavior of the images. 
The reason for this is illustrated in Fig.~\ref{fig1} for a single spin system. 
Three neighboring images, $\nu-1$, $\nu$ and $\nu+1$ are placed equidistantly along a geodesic 
path which lies in the plane of the figure. 
The forward-difference approximation is used to define the local tangent to the path at image 
$\nu$: $\hat{\tau}^\nu=\left(\vec{m}^{\nu+1}-\vec{m}^\nu\right)/\left|\vec{m}^{\nu+1}-\vec{m}^\nu\right|$. 
Since the distance between adjacent images is the same in this case, the spring force is zero. 
Even if the transverse component of the true force is calculated 
by subtracting out the component parallel to the NEB tangent $\hat{\tau}^\nu$ 
and projecting on the tangent space so as to satisfy the magnetic constraint, 
the resulting force still has a component along the path. 
The true force then affects the distribution of images along the path. 
This will cause image $\nu$ to move towards the lower energy region of the path, i.e. slide down. 
Control of the arrangement images along the path is then lost and, 
as a result, the resolution of the path in the critical regions becomes poor. 
The value chosen for the spring constant then also becomes an important issue.
This behavior is demonstrated in an application to a single spin test problem in the following section. 
The problem can be avoided by projecting the tangent on the tangent space as illustrated in Fig. 1a.

An initial path is needed to start the GNEB calculation.
In many cases it is sufficient to place the images initially along the geodesic path between $\bm{M}^I$ and $\bm{M}^F$. 
This is analogous to the straight-line interpolation between the initial and final point 
often used to start NEB calculations of atomic rearrangements. A simple way to obtain the intermediate images 
$\bm{M}^\nu=\left(\vec{m}_1^\nu,\vec{m}_2^\nu,\ldots,\vec{m}_N^\nu\right)$ along the geodesic path is to rotate each magnetic momentum from the initial orientation to the final one using Rodrigues' rotation formula:
\begin{equation}
\label{eq:rodrig}
\vec{m}_i^\nu = \vec{m}_i^I\cos\omega_i^\nu+(\vec{k}_i\times\vec{m}_i^I)\sin\omega_i^\nu.
\end{equation}
Here $\vec{m}_i^\nu$ is the $i$-th unit magnetic vector in image $\nu$th and 
$\omega_i^\nu=(\nu-1)\Delta\omega_i$ is an angle of rotation. 
If $Q$ images are used in the chain, then $\Delta\omega_i = \omega_i/(Q-1)$, 
where $\omega_i$ is an angle between vectors $\vec{m}_i^I$ and $\vec{m}_i^F$. 
A unit vector $\vec{k}_i$ describing an axis of rotation is defined as:
\begin{equation}
\label{axis}
\vec{k}_i=\frac{(\vec{m}_i^I\times\vec{m}_i^F)}{\bigl|(\vec{m}_i^I\times\vec{m}_i^F)\bigr|}.
\end{equation}
As with the NEB for atomic rearrangements, when multiple MEPs are present, 
the optimization procedure will most likely lead to convergence to the MEP 
closest to the initial path and some sampling of the various MEPs is needed to find the optimal one, or -- more generally -- 
all the relevant ones. 

A summary of the GNEB algorithm is given in Appendix~\ref{app:gneb_algo}.
A climbing image can be used in a GNEB calculation, a CI-GNEB method, analogous to the CI-NEB method described in the previous section. The highest energy image is chosen to be the climbing image and the force 
on it is only the inverted parallel component of the true force without the addition of a spring force
\begin{equation}
\label{eq:forceCI-GNEB}
\bm{F}^{\nu_{CI}}_{GNEB} = \mathcal{P}_\mathcal{T}(-\nabla E\left(\bm{M}^{\nu_{CI}}\right)+
2\bigl(\nabla E\left(\bm{M}^{\nu_{CI}}\right)\cdot\hat{\bm{\tau}}^{\nu_{CI}}_\mathcal{T}\bigr)\hat{\bm{\tau}}^{\nu_{CI}}_\mathcal{T}).
\end{equation}
The CI-GNEB method is most useful when the energy barrier is narrow and sharp compared with the length of the path.

As for NEB calculations of atomic rearrangements, it is important to interpolate the energy along GNEB paths using information 
contained in the force along the path as well as the energy at each image, as described in Appendix~\ref{app:interpol}.  
This is important for obtaining an accurate estimate of the saddle point energy as well as to identify possible intermediate minima. 

The formalism described above for the (CI-)GNEB method is independent of the coordinate system. 
We have chosen here to use spherical polar coordinates since they naturally fulfill the magnetic constraints. 
A configuration of a system of $N$ spins is specified by the coordinates 
$(\theta_1,\phi_1,\theta_2,\phi_2, \dots \theta_N,\phi_N )$
and all displacements are then automatically within the $\mathcal{R}$-manifold.


\section{Results}
\subsection{Test problems}

The GNEB method is first illustrated with calculations of the MEP of a transition involving a single spin
where the energy surface can be visualized easily. 
The energy is given by a sum of Gaussian functions of geodesic distances, as described in Appendix~\ref{app:test_surf}.

First, several minimization calculations were carried out where parameters of the energy surface and starting points were chosen randomly. The minimization algorithm is based on an equation of motion where the
velocity is projected in such a way as to converge on a minimum, as described in Appendix~\ref{app:vpo}.
The test calculations were done to assess the robustness of the minimization algorithm, especially near the poles.
In some of the calculations, the energy surface was rotated so that one of the minima conicided with a pole. 
With the special treatment of the polar regions, described in Appendix~\ref{app:vpo},
the minimization converged witout problems. 
Furthermore, the number of iterations needed to reach convergence did not tend to be larger 
when the minimum coincided with a pole. 

\begin{figure}[h!]
\centering
\includegraphics[width=0.4\columnwidth]{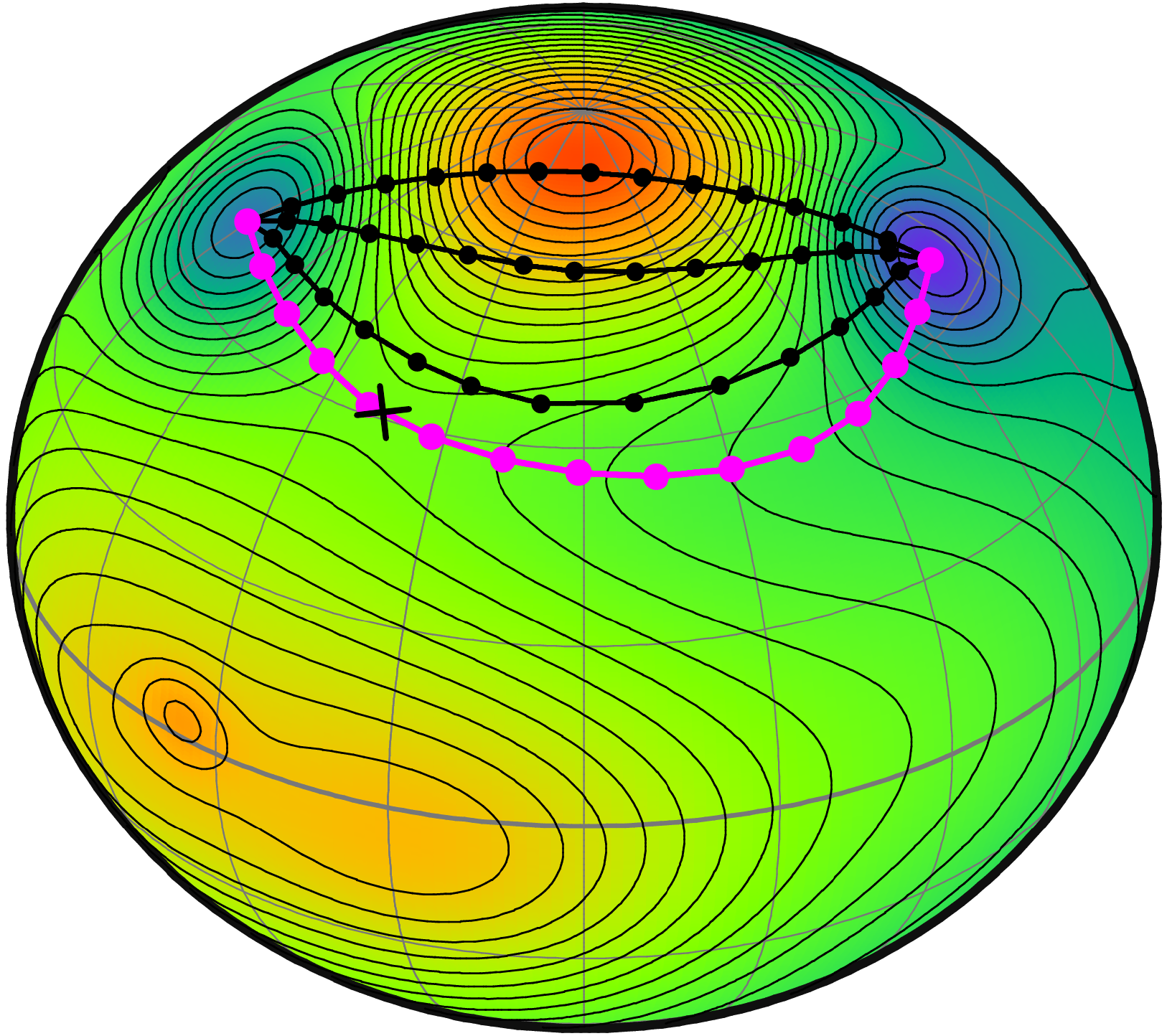}
\caption{
A GNEB calculation of an MEP in a single spin system. The positions of the images are shown with filled circles. 
The initial path was chosen to lie 
along the short geodesic path between the energy minima (indicated by blue color) 
and turned out to lie near a maximum (indicated by red color). 
The converged path is shown in pink. 
Two intermediate configurations of the path are also shown. The saddle point is marked with a cross.
$Q=15$ images were used in the calculation. 
}
\label{fig2}
\end{figure}

Fig.~\ref{fig2} shows a GNEB calculation of the MEP for a transition of the spin from one minimum to another. 
The initial distribution of images was generated along the short geodesic path connecting the minima. The geodesic path is obviously not an optimal one since it lies close to an energy maximum. 
The images where then iteratively brought to the MEP using the GNEB method with the minimization algorithm described in Appendix~\ref{app:vpo}.

\subsubsection{Effect of not projecting the tangent on the tangent space of $\mathcal{R}$}

\begin{figure}[h!]
\centering
\includegraphics[width=0.35\columnwidth]{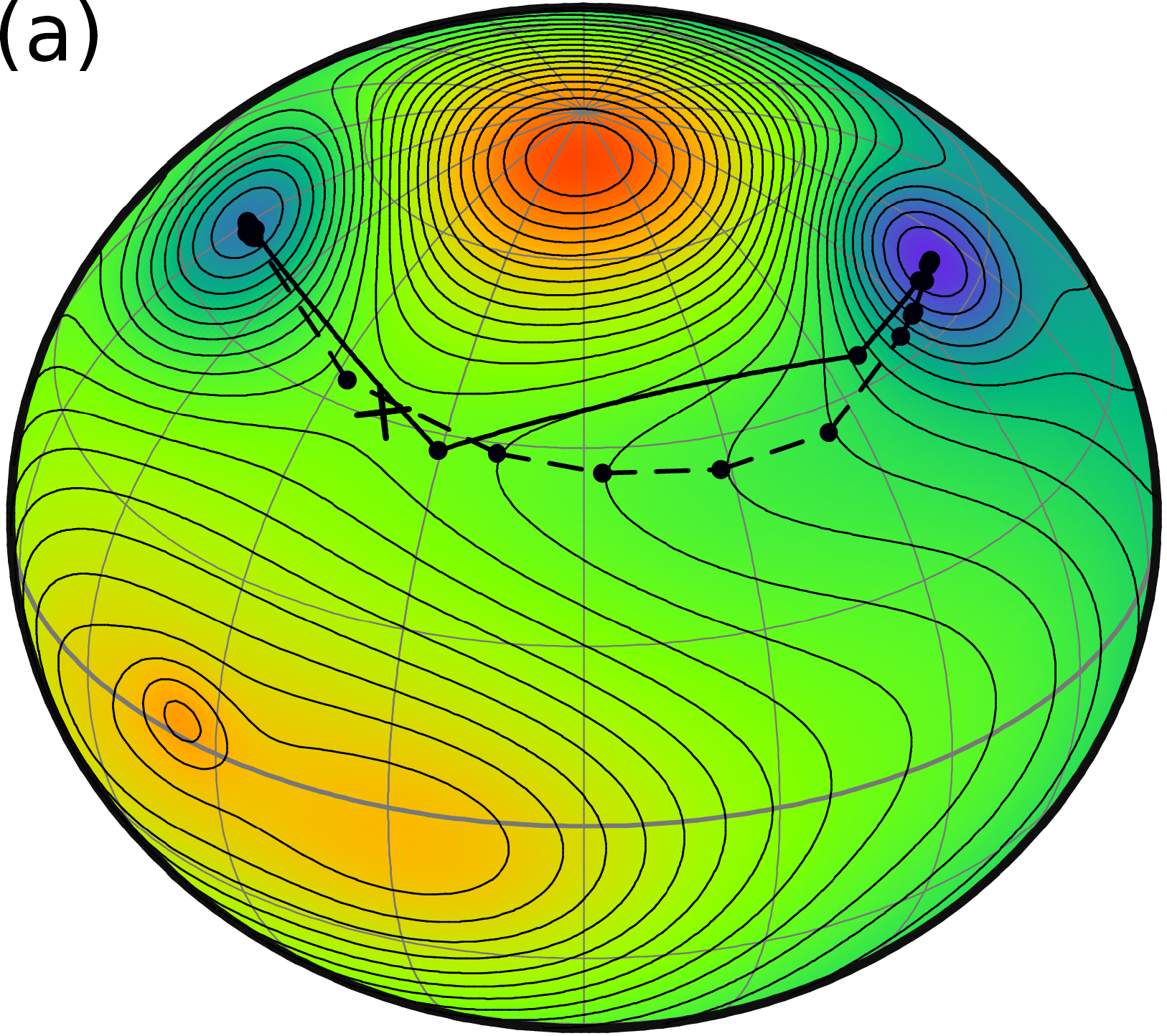}
\hspace{2cm}
\includegraphics[width=0.35\columnwidth]{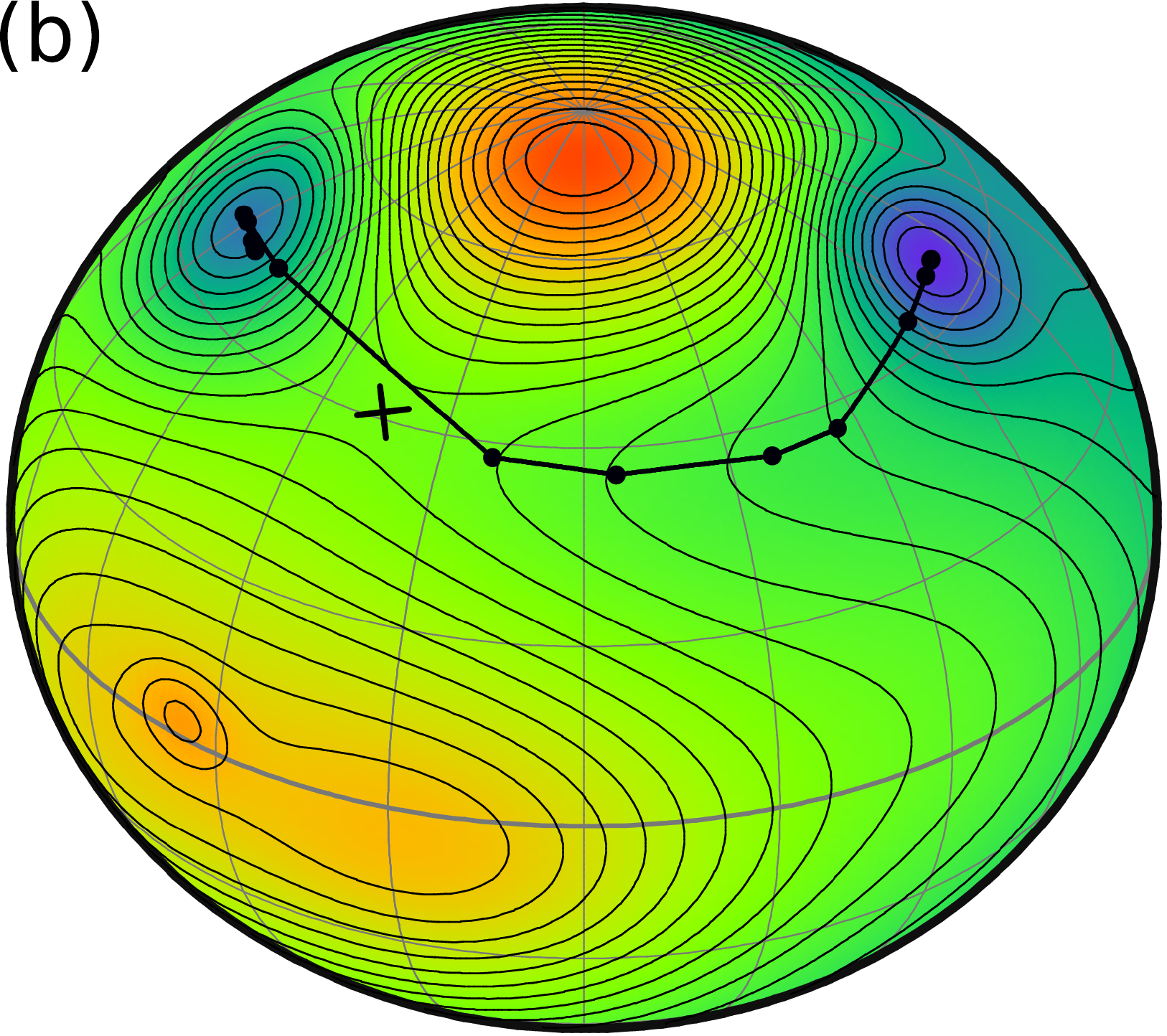}
\caption{
Same test problem as in Fig.~\ref{fig2}. The images are shown with filled circles and the saddle point with a cross.
(a) 
Results of calculations where the tangent is not projected on the tangent space. The initial distribution of the images was chosen to be along the geodesic path connecting the energy minima. The path after 200 time steps 
is shown with a dashed line and the path after 400 steps is shown with a solid line. 
The parallel component of the true force is not completely projected out
in this case and that makes the images slide down from the barrier region. 
(b) 
Results of optimization without springs but including the full GNEB force projections. 
Because of kinks on the path, a wiggling motion during the iterative minimization enables the images to 
gradually slide down in energy. As a result, the resolution near the saddle point becomes poor. 
Small random displacements were added to the initial arrangement of the images along the geodesic. 
}
\label{fig3}
\end{figure}

As illustrated in Fig.~\ref{fig1}, 
the true force and the spring force are not decoupled properly unless the tangent to the path is projected on the tangent space. 
Without this projection, interference between the forces lead to loss of control of the images, 
such as sliding down from the saddle point region towards the minima. 
Fig.~\ref{fig3}(a) shows the results of such a calculation. 
After 200 iterations the images have moved to the vicinity of the MEP, but the spacing between images is uneven and 
the resolution of the path in the barrier region is unsatisfactory. 
Since the parallel component of the true force is not projected out completely, the images slowly slide down from the barrier region. After 400 steps most of the images have slid down into the potential wells and the information about the saddle point region is lost. 
It is still possible to obtain a reasonable estimate of the MEP without the projection on the tangent space
but it requires choosing just the right value for the spring constant. 
Such behavior has indeed been reported before~\cite{berkov_07}. 

When the tangent to the path is correctly defined by including the projection on the tangent space, the method is, 
however, stable and quite insensitive to the choice of the value of the spring constant.
In fact, the GNEB calculations for the test problem in Fig.~\ref{fig2} were found to give nearly identical results as the
value of the spring constant was varied over five orders of magnitude.
For complex transitions involving multiple spins and rotation in 3-dimensions, 
proper convergence to the MEP within a 
tight force tolerance (such as the 10$^{-9}~$meV/radian tolerance specified in the present calculations) 
may not occur for any value of the spring constant unless the tangent to the path is projected on the tangent space.

\subsubsection{Effect of skipping the springs}

Since the forces are decoupled in the GNEB method, the value of the spring constant is not critical. For some simple systems, 
reasonable results can even be obtained with the NEB without the inclusion of springs~\cite{dittrich_02}, 
but control of the distribution of the images is then lost and the minimization does not converge properly. 
Fig.~\ref{fig3}(b) shows the results for the same test problem as the one shown in Fig.~\ref{fig2}. 
The images in the initial path were arranged along the geodesic between the two minima 
with small random displacements added before starting a GNEB calculation without springs. 
The calculation converges in this case, i.e. perpendicular component of the true force drops below the set 
tolerance, but the final distribution of the images is uneven, as most of the images have slid down towards the energy minima. 
Without the springs, the method becomes unstable with respect to small perturbations. 
The inclusion of the springs as in Fig.~\ref{fig2} ensures convergence to the MEP with an equal distribution of images 
when the same value of the spring constant is chosen for all the springs.


\subsection{Annihilation of a skyrmion}

The GNEB method was used to calculate the MEP for annihilation of a skyrmion to form a homogeneous, collinear 
state.
While these are topologically distinct magnetic states, there is a finite energy barrier for transitions between the two.
The MEP not only gives an estimate of the activation energy barrier but also reveals 
the microscopic mechanism of the skyrmion annihilation, or -- in reverse -- skyrmion formation.

\begin{figure}[h!]
\centering
\includegraphics[width=0.7\columnwidth]{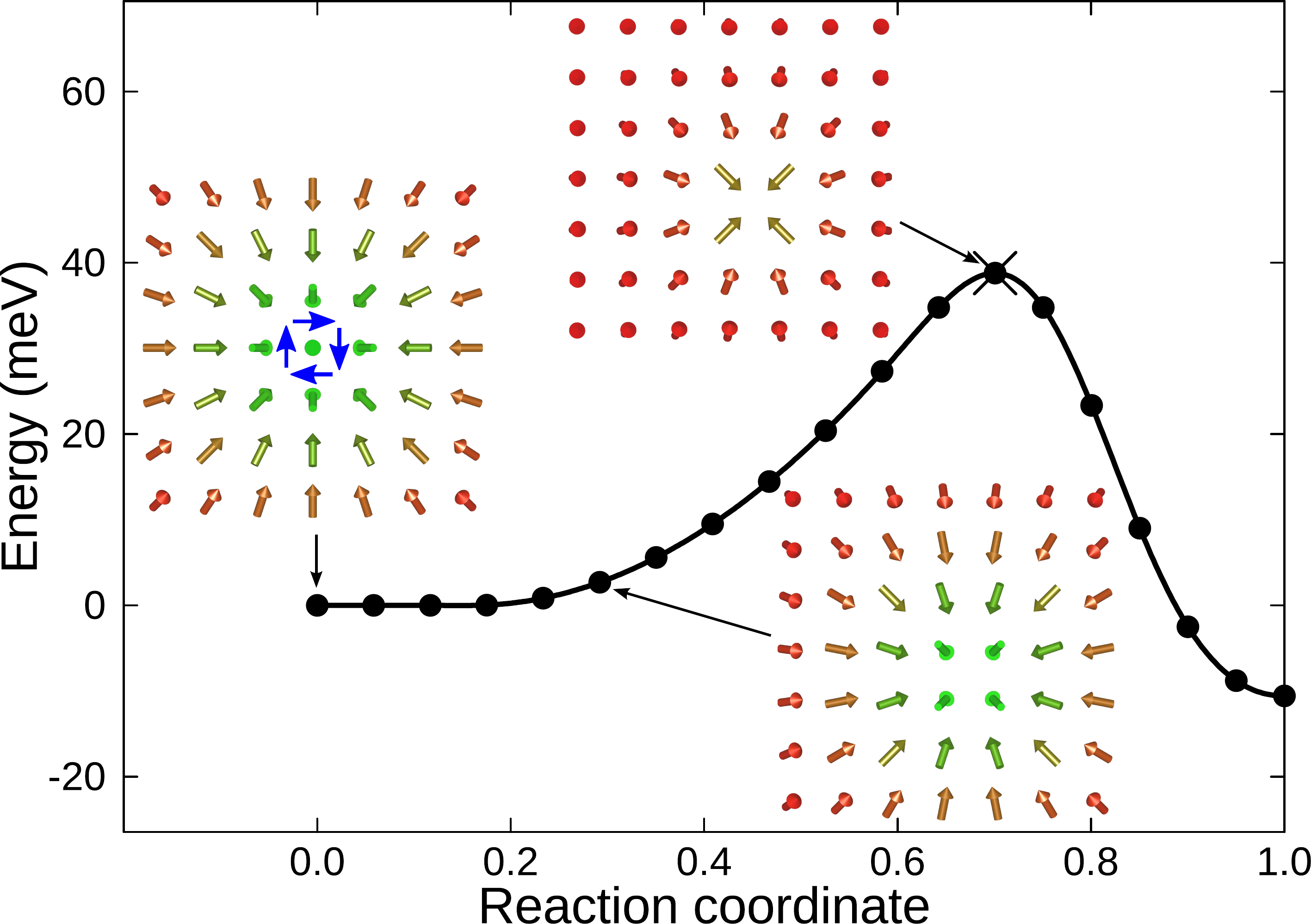}
\caption{
Minimum energy path showing the mechanism and activation energy for the annihilation of an isolated skyrmion. 
The filled circles indicate images along the converged path obtained using the CI-GNEB method.  
The interpolation method described in Appendix~\ref{app:interpol} was used.  
Insets show the orientation of magnetic moments at three configurations. 
The color indicates the value of the z-component of the magnetic vectors 
(red $\leftrightarrow$ up, green $\leftrightarrow$ down).
Only a part of the simulated system is shown. 
The ground state corresponds to collinear ordering of the magnetic vectors. 
The thin, blue arrows in the left-most inset show the direction of the pseudovectors for the DM interaction of a central 
moment with its four nearest neighbors. 
}
\label{fig4}
\end{figure}

The emergence of magnetic skyrmions is the result of an interplay of multiple interactions, 
where the Dzyaloshinskii-Moriya (DM) interaction and 
noncollinear isotropic exchange are believed to play a crucial role~\cite{kiselev_11,fert_13}. 
A discussion of the microscopic origin of skyrmion states is, however, outside the scope of this article. 
Here, skyrmion configurations are obtained using a simplified phenomenological 
model based on a Heisenberg-type Hamiltonian including isotropic exchange, 
DM interaction as well as interaction with an external field. 
The energy of the system is defined as:
\begin{equation}
\label{eq:skyrm}
E = -\frac{1}{2}J\sum_{\left<i,j\right>}^N \vec{m}_i\cdot\vec{m}_j-\frac{1}{2}\sum_{\left<i,j\right>}^N \vec{D}_{ij}\cdot[\vec{m}_i\times\vec{m}_j]-\vec{B}\sum_{i}^N \vec{M}_i
\end{equation}
Here, $J$ is the isotropic exchange parameter, the $\vec{D}_{ij}$ are the DM vectors and 
$\vec{B}$ denotes an external magnetic field. 
Values of the parameters of the Hamiltonian were taken from Ref.~\cite{kiselev_15}. 
The exchange interaction is included only between nearest neighbors (indicated by angular brackets). 
The spins are located in the XY-plane at the nodes of a 2-dimensional square lattice. 
The size of the simulated domain was $21\times21$ spins, which is significantly larger than the 
size of the skyrmion. Rigid boundary conditions were applied, essentially mimmicking an infinite 2-dimensional layer.
The magnetic field was applied along the normal to the plane. The DM vectors lie in the plane as shown in Fig.~\ref{fig4}.

Minimization calculations starting from random initial configurations of the spins revealed two stable states in the system. 
The global energy minimum represents a collinear state with all the magnetic vectors aligned along the field $\vec{B}$. 
A noncollinear, metastable state corresponding to an isolated skyrmion was also found.
The skyrmion center coincides with a lattice site, see Fig.~\ref{fig4}.
The CI-GNEB calculation was started by generating 19 images along the geodesic path between these minimum energy
configurations and then applying the minimization algorithm described in Appendix~\ref{app:vpo}. 
until the force on each magnetic momentum had dropped below $10^{-9}~$meV/radian. 
The calculation converged to the MEP without any problem. 
Fig.~\ref{fig4} shows the result.
The MEP reveals a mechanism for the annihilation of the skyrmion. At first, the skyrmion moves as a whole without changing its size and shape, shifting its center in between the atomic sites. 
This shift involves almost no change in energy. After that, the spins rotate symmetrically 
causing the skyrmion to gradually shrink and eventually disappear.
This is a challenging calculation involving out-of-plane rotation of the magnetic momenta, 
as can be seen from the insets of Fig.~\ref{fig4}.

The activation energy for the annihilation of the skyrmion state was calculated to be 40 meV, 
but 50 meV for its formation from the collinear state. 
The saddle point configuration is shown in an inset of Fig.~\ref{fig4}. There, the four central spins lie in the XY-plane and point towards each other. Each of them is perpendicular to the nearest neighbors but antiparallel to next nearest neighbors. 


\subsection{Annihilation of an antivortex}

The noncollinar Alexander-Anderson (NCAA) model has been found to describe well magnetism of 
itinerant electrons in 3d transition metals~\cite{uzdin_12,uzdin_09a}. 
While the evaluation of the energy requires self-consistency calculations,  
analytical forces provided by a force theorem~\cite{bessarab_14} make it relatively easy 
to navigate on the energy surface and find local minima corresponding to (meta)stable magnetic states with possibly 
complex, non-collinear ordering of the magnetic moments. 
Calculations of a 7x7 atomic row island of Fe atoms on a substrate have been shown to lead to an antivortex state when
the parameter values are slightly different from the optimal values of an Fe crystal
under ambient conditions~\cite{bessarab_14}.
Such a slight change in the model parameters could be the result of an external perturbation such as an external 
electrical field or the presence of impurities or defects.
The metastable antivortex state was found by starting from a random initial 
orientation of the magnetic moments and then minimizing the energy to bring the system to a minimum energy configuration.  
This state can be described as a symmetrical, saddlelike arrangement of the magnetic moments
in the center of the island. The total in-plane magnetization is zero, 
while the out-of-plane magnetization is non-zero mainly due to four magnetic moments 
near the center of the island pointing out of plane.

\begin{figure}[h!]
\centering
\includegraphics[width=0.7\columnwidth]{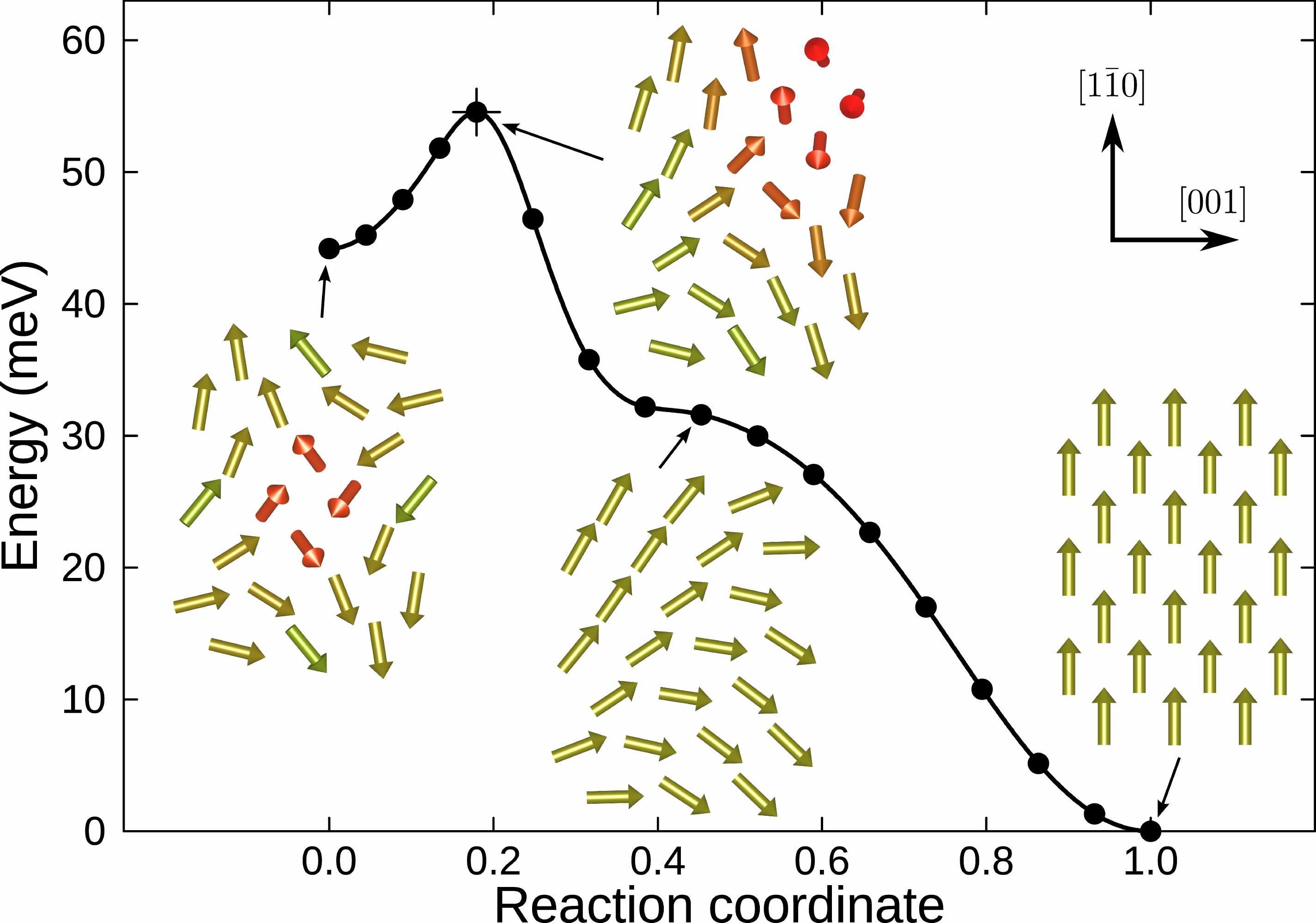}
\caption{
Minimum energy path showing the mechanism and activation energy for the annihilation of an antivortex state obtained
in a noncollinear Alexander-Anderson model of a 7x7 island of Fe atoms. 
The filled circles indicate images along the converged path obtained using the CI-GNEB method. 
Insets show the orientation of the magnetic moments at various points along the path, as indicated by arrows,
colored with the same scheme as in Fig.~\ref{fig4}.
The ground state corresponds to collinear ordering of the magnetic vectors. 
The interpolation method described in Appendix~\ref{app:interpol} was used. 
}
\label{fig5}
\end{figure}

The CI-GNEB method was used to calculate the MEP for the transition between the antivortex state 
and the homogeneous ground state,
and the results are shown in Fig.~\ref{fig5}.  
The calculation was considered converged when the force on each magnetic momentum 
had dropped below $10^{-9}~$meV/radian.
From these results, the activation energy for the annihilation 
as well as the formation of the antivortex state can be obtained. 
As the system progresses along the MEP, the saddle-like excitation moves along the diagonal of the island towards 
one of the corners (the upper right hand corner in the insets of Fig.~\ref{fig5}) where it leaves the island. 
As with the skyrmion annihilation, finding this MEP is a challenging calculation involving out-of-plane rotation of the 
magnetic momentum vectors, as can be seen from the insets of Fig.~\ref{fig5}.
The maximum energy configuration was verified as a SP configuration 
using the method described in Appendix~\ref{app:verif_sp}.
The activation energy for the annihilation of the antivortex state was calculated to be 10 meV, 
but 55 meV for its formation from the collinear state. 
The pre-exponential factor in the rate constant was determined using HTST and found to be 
1.4$\times$10$^{13}$ s$^{-1}$ for the annihilation of the antivortex.  
For the reverse transition, the pre-exponential factor was larger,
2.6$\times$10$^{13}$ s$^{-1}$, showing that the vibrational entropy of the antivortex state is larger than that
of the homogeneous state.


\section{Discussion}

The results presented here on the annihilation of the skyrmion and antivortex states are preliminary 
in that they have been obtained for only particular systems and Hamiltonians. They are presented here mainly to 
illustrate possible transition mechanisms and give an indication of the thermal stability in these simple systems.
It is not known at this time whether similar transition mechanism will be obtained for other models supporting 
these noncollinear states. This remains to be seen.

The existence of a local minimum on the energy surface of the NCAA model corresponding to an antivortex state 
demonstrates the complex exchange interactions included in the model even though it contains only a few parameters.
Such a behavior cannot be obtained within a Heisenberg-type model unless additional phenomenological terms and
additional parameters are introduced in the Hamiltonian. 
In the NCAA model, complex noncollinear states appear quite naturally.
It may be possible to obtain skyrmion states within the NCAA model. Again, this remains to be seen.

Most importantly, the calculations of the annihilation of the skyrmion and antivortex states presented here 
illustrate the applicability of the GNEB method to complex magnetic transitions.  
While the NEB method has been used already in several studies of magnetic transitions
~\cite{dittrich_02,dittrich_03,thiaville_03,e_03,dittrich_04,dittrich_05,suess_06,suess_07,berkov_07, goll_07,krone_10,
visscher_12,fiedler_12,bessarab_12,tudosa_12,bessarab_13,bessarab_14}
convergence problems and sensitivity to parameters such as the spring constant have been noted and in light of the analysis
presented here it is likely that rigorous convergence, defined here as a drop in magentic forces below $10^{-9}~$meV/radian,
was in some cases not obtained. The extension of the NEB method presented and used here, the GNEB method, 
resolves these convergence issues and provides a tool for the study of more complex magnetic 
systems than could be studied with the NEB method. 

The GNEB method assumes a certain shape of the configuration space: 
a direct product of 2D spheres associated with each magnetic moment in the system.
This requirement is clearly fulfilled by the quasi-classical Heisenberg-type models and micromagnetics, where the length of magnetic vectors is usually fixed and does not depend on orientation of the spins. 
It is also fulfilled in the quantum-mechanical treatment of the system within an adiabatic approximation, where the 
magnitude of the magnetic 
moments is a function of the orientation~\cite{antropov_96, bessarab_14}. The magnitude of the magnetic moments 
is assumed to adjust instantaneously to changes in orientation. The energy surface is entirely defined by the
orientation of the magnetic momentum vectors. 
In general the magnetic force can 
contain a term proportional to the derivative of the energy with respect to the length of magnetic moments, 
which accounts for the change in magnitude of magnetic moments as magnetic vectors rotate. 
However, this term vanishes for all important cases such as density functional theory or NCAA model calculations
as shown by a magnetic force theorem \cite{liechtenstein_87, bessarab_14}.

Here we present some practical information which we find useful for a successful GNEB calculation.
If interpolation of the energy along the path using the method described in Appendix~\ref{app:interpol} gives an 
indication of an intermediate minimum, it is recommended to divide the path up into segments and
study transitions between each pair of adjacent minima separately. 
The location of the minimum should first be carefully identified by carrying out an energy 
minimization and then calculate the MEPs between each pair of adjacent minima in separate GNEB calculations. 
The full MEP is, thereby, obtained as a composition of segments of MEPs, each passing through only one SP.

The question arises when to use GNEB and when to use CI-GNEB.
The latter is particularly efficient when the energy maximum along the path is well defined. 
However, there are important 
exampes of magnetic transitions where the barriers are flat, making it hard to identify the highest energy image. 
Flat barriers appear, for example, when the mechanism of a magnetic transition involves 
domain wall propagation without significant change in energy~\cite{braun_12,bessarab_13, dittrich_02}. 
In such cases, it is better to use the regular GNEB method rather than the CI-GNEB method.
A good strategy for finding MEPs is, therefore, to start the calculation using the GNEB method and 
inspect the energy along the path after a few tens of iterations or when the GNEB forces have dropped below 
some predefined value (significantly larger than $10^{-9}~$meV/radian). 
If there is an indication of a flat barrier, then continue using GNEB until convergence (see Appendix~\ref{app:gneb_algo}). 
If the path seems to have a well-defined maximum, then switch to the CI-GNEB method. 
During the CI-GNEB calculations, the energy of the climbing image might become lower than 
that of some other image in the chain. In this case, the climbing image should to be reassigned.

The GNEB method does not imply a particular optimization method. 
Any optimization method based only on the evaluation of the forces, 
i.e. the first derivatives of the energy with respect to the configuration parameters, can be used. 
The objective function corresponding to the GNEB forces is not well defined, however, because of the force projections and
magnetic constraints.  
It remains to identify the most efficient optimizers for the GNEB method, similar to what has been done 
for the NEB method~\cite{sheppard_08}. The method presented in Appedix D is stable and reliable, but probably not optimal in 
terms of efficiency. It is likely that a switch to some quadratically convergent optimization method is beneficial 
when the images are relatively close to the MEP, while the use of a more stable method, such as the one presented in Appedix D, 
is safer when starting from the arrangement of the images along the geodesic path. Since the most important point along the 
path is the highest energy SP, it can be efficient to converge the MEP only to a rather high tolerance but then converge
with a low tolerance on the SP using, for example, the minimum mode following method~\cite{henkelman_99,olsen_04}.

To summarize, we emphasize that the GNEB method differs from the NEB method in that 
the curvature of the configuration space of magnetic systems is taken into account. 
A particularly important 
aspect of the GNEB method is the projection of the tangent to the path on the tangent space. 
If this projection is not implemented, a rigorous convergence of to the MEP will likely not be obtained. In fact, 
we tried to reproduce the results obtained with the (CI-)GNEB calculations on the skyrmion and antivortex 
annihilation using simpler implementations of the NEB method for magnetic transitions where projection of the 
tangent to the path on the tangent space is not included (as in refs.~\cite{dittrich_02,suess_07}), 
but could not reach convergence to the MEP for any value of the spring constant.


\section{Acknowledgements}
We thank N.S. Kiselev for helpful discussions on skyrmions and for providing information on the model Hamiltonian. 
This work was supported by the Russian Foundation of Basic Research (Grants No. 14-02-00102, and No. 14-22-01113 ofi-m,), the Icelandic Research Fund and the Nordic-Russian Training Network for Magnetic Nanotechnology (NCM-RU10121). 
PB gratefully acknowledges support from the G\"oran Gustafsson Foundation and F. S. Bessarab for useful discussions.


\appendix


\section{Evaluation of path tangent}
\label{app:path_tang}

The tangent to the path at image $\hat{\bm{\tau}}^\nu$
can be estimated using the coordinates of the two adjacent images in a straightforward way as 
\begin{equation}
\hat{\bm{\tau}}^\nu = \frac{\bm{M}^{\nu+1}-\bm{M}^{\nu-1}}{\left|\bm{M}^{\nu+1}-\bm{M}^{\nu-1}\right|}
\end{equation}
but this can lead to the formation of kinks in the path and, as a result, cause instabilities that slow down or even prevent convergence to the 
MEP~\cite{NEBleri}. 
This problem can be reduced by using a better definition of the tangent based on either forward or backward difference depending on the energy 
of the image~\cite{newtang}: 
\begin{equation}
\bm{\tau}^\nu = \begin{cases}
\bm{\tau}^\nu_+,&\text{if $E^{\nu+1}>E^{\nu}>E^{\nu-1}$,}\\
\bm{\tau}^\nu_-,&\text{if $E^{\nu+1}<E^{\nu}<E^{\nu-1}$,}\\
\end{cases}
\end{equation}
where $\bm{\tau}^\nu_+=\bm{M}^{\nu+1}-\bm{M}^{\nu}$, and $\bm{\tau}^\nu_-=\bm{M}^{\nu}-\bm{M}^{\nu-1}$ and $E^{\nu}=E\left(\bm{M}^\nu\right)$. 
If image $\nu$ is close to an energy minimum ($E^{\nu+1}>E^{\nu}<E^{\nu-1}$) or to an energy maximum ($E^{\nu+1}<E^{\nu}>E^{\nu-1}$) along the path, the tangent is taken to be a weighted average of $\bm{\tau}^\nu_+$ and $\bm{\tau}^\nu_-$ so as to ensure a smooth switch between the forward and the backward difference schemes:
\begin{equation}
\bm{\tau}^\nu = \begin{cases}
\bm{\tau}^\nu_+\Delta E^\nu_{max}+\bm{\tau}^\nu_-\Delta E^\nu_{min},&\text{if $E^{\nu+1}>E^{\nu-1}$,}\\
\bm{\tau}^\nu_+\Delta E^\nu_{min}+\bm{\tau}^\nu_-\Delta E^\nu_{max},&\text{if $E^{\nu+1}<E^{\nu-1}$,}\\
\end{cases}
\end{equation}
where $\Delta E^\nu_{max}=\max\bigl(\left|E^{\nu+1}-E^{\nu}\right|,\left|E^{\nu-1}-E^{\nu}\right|\bigr)$, and $\Delta E^\nu_{min}=\min\bigl(\left|E^{\nu+1}-E^{\nu}\right|,\left|E^{\nu-1}-E^{\nu}\right|\bigr)$. 



\section{Evaluation of the geodesic distance}
\label{app:geo_dist}

An essential aspect of the GNEB method is to evaluate the geodesic distance between two locations of a magnetic vector. 
Three different equations for this calculation are given here. The formulas are mathematically equivalent but they differ in numerical stability.  

The spherical law of cosines~\cite{korn_00} 
\begin{equation}
l^{\nu,\mu}_i = \arccos(\vec{m}_i^\nu\cdot\vec{m}_i^\mu).
\end{equation}
is the simplest formula but it can lead to significant round-off errors when the distance 
between the two points on the unit sphere is small.

The haversine formula works well for small as well as large distances but is more complicated~\cite{sinnott_84}
\begin{equation}
l^{\nu,\mu}_i = 2\arcsin\sqrt{\frac{1-\vec{m}_i^\nu\cdot\vec{m}_i^\mu}{2}}=
2\arcsin\sqrt{\sin^2\left(\dfrac{\theta_i^\nu-\theta_i^\mu}{2}\right)+\sin\theta_i^\nu\sin\theta_i^\mu\sin^2\left(\dfrac{\phi_i^\nu-\phi_i^\mu}{2}\right)}.
\end{equation}
Here, $\theta_i^\nu$ and $\phi_i^\nu$ are polar and azimuthal angles defining orientation of $\vec{m}_i^\nu$.
The only problem arises for antipodal points (on directly opposite sides of the sphere), where the formula suffers from round-off errors.

Vincenty's formula is~\cite{vincenty_75} 
\begin{equation}
l^{\nu,\mu}_i = \text{arctan2}\bigl(\left|\vec{m}_i^\nu\times\vec{m}_i^\mu\right|,\vec{m}_i^\nu\cdot\vec{m}_i^\mu\bigr).
\end{equation}
where the function $\text{arctan2}(y,x)$ is defined as follows:
\begin{equation}
\text{arctan2}(y,x)=
\begin{cases}
\arctan \dfrac{y}{x}, &\text{if $x>0$}\\
\arctan \dfrac{y}{x}+\pi, &\text{if $y\ge0$, $x<0$}\\
\arctan \dfrac{y}{x}-\pi, &\text{if $y<0$, $x<0$}\\
+\frac{\pi}{2}, &\text{if $y>0$, $x=0$}\\
-\frac{\pi}{2}, &\text{if $y<0$, $x=0$}\\
\text{undefined}, &\text{if $y=0$, $x=0$}.
\end{cases}
\end{equation}
It works well for all caes but is the most complicated expression and requires the definition of the $\text{arctan2}$ function 
because the usual $\arctan$ function gives values in between $-\pi/2$ and $\pi/2$ while the geodesic length on the unit sphere should be between $0$ and $\pi$.  We have used Vincenty's formula in the calculations presented here.


\section{Summary of GNEB algorithm}
\label{app:gneb_algo}

A brief, schematic summary of a GNEB calculation is as follows:

1. Identify two minima on the energy surface correspondig to the initial and final configurations, 
$\bm{M}^I=\left(\vec{m}_1^I,\vec{m}_2^I,\ldots,\vec{m}_N^I\right)$ and $\bm{M}^F=\left(\vec{m}_1^F,\vec{m}_2^F,\ldots,\vec{m}_N^F\right)$, 
between which an MEP is to be found.

2. Create an initial path by constructing a chain of intermediate images between $\bm{M}^I$ and $\bm{M}^F$. 
When the geodesic path is used to construct the initial path, 
an intermediate image, $\nu$, with coordinates 
$\bm{M}^\nu=\left(\vec{m}_1^\nu,\vec{m}_2^\nu,\ldots,\vec{m}_N^\nu\right)$ 
is generated using Eqs.(\ref{eq:rodrig}) and (\ref{axis}).

3. For each image, set the velocity, $\bm{\EuScript{V}}^\nu$, to zero.

\noindent
This completes the initialization of the iterative algorithm.

\vskip 0.3 true cm

4. Calculate the energy of the images, $E\left(\bm{M}^\nu\right)$.

5. For each image, calculate the tangent to the path, as described in Appendix~\ref{app:path_tang}. 
Then project the tangent on the tangent space, Eqs.(\ref{eq:TanProj}).

6. For each image, calculate the transverse component of the true force, Eq.(\ref{eq:Gperp}).

7. Calculate the geodesic distance between pairs of adjacent images, Eq.(\ref{eq:geodesicdist}).

8. For each image, calculate the parallel component of the spring force, Eq.(\ref{eq:GFsparallel}).

9. For each image, calculate the total force by adding the transverse component of the true force projected on the tangent space to 
the parallel component of the spring force, Eq.(\ref{eq:forceGNEB}).

10. If the GNEB force on any one of the magnetic vectors in any one of the images is larger than the tolerance, $F_{tol}$, then 
update the position and velocity according to the optimization algorithm in Appendix~\ref{app:vpo} 
and go back to step 4. 
Otherwise stop, and interpolate the energy along the path using the method in Appendix~\ref{app:interpol} and verify that the highest energy point is a first
order saddle point using the approach in Appendix~\ref{app:verif_sp}.


\section{Interpolation between images}
\label{app:interpol}

The final, relaxed configuration of the images obtained from a GNEB calculation gives a discrete representation of the MEP. 
Except for some simple, highly symmetrical cases, the highest energy image in the discrete representation of the path
will likely not coincide with the energy maximum of the MEP, i.e.  a SP on the energy surface. 
In most cases, the SP lies in between the images. An accurate interpolation of the path between the images is, therefore, important.
It is also useful for identifying intermediate minima along the path.
Not only the energy at each image is available for the interpolation but also the component of the energy gradient along the path at each image, 
$\delta^\nu = \left(\nabla E\left(\bm{M}^\nu\right)\cdot\hat{\bm{\tau}}^\nu_\mathcal{T}\right)$. 
This input can be used to construct a piecewise cubic polynomial interpolation. The energy of the system is approximated as 
\begin{equation}
\label{eq:interpol}
E(\lambda) = a_\nu(\lambda-\lambda_\nu)^3+b_\nu(\lambda-\lambda_\nu)^2+c_\nu(\lambda-\lambda_\nu)+d_\nu,\hspace{0.5cm}\lambda\in\left[\lambda_\nu;\lambda_{\nu+1}\right]
\end{equation}
where $\lambda$ is the displacement along the poly-geodesic MEP and $\lambda_\nu$ is given by
\begin{equation}
\label{eq:lambda}
\lambda_\nu=
\begin{cases}
0, &\text{if $\nu=1$,}\\
\lambda_{\nu-1}+L(\bm{M}^{\nu-1},\bm{M}^{\nu}), &\text{if $\nu=2,\ldots,Q$.}
\end{cases}
\end{equation}
For each segment, the $\left[\lambda_\nu;\lambda_{\nu+1}\right]$, parameters $a_\nu$, $b_\nu$, $c_\nu$ and $d_\nu$ 
are defined so as to enforce continuity of the energy and its derivative at both ends of the segment:
\begin{align}
a_\nu &= \frac{\delta^{\nu+1}+\delta^\nu}{\left(\lambda_{\nu+1}+\lambda_\nu\right)^2}-\frac{2\left(E^{\nu+1}-E^\nu\right)}{\left(\lambda_{\nu+1}+\lambda_\nu\right)^3},\\
b_\nu &= -\frac{\delta^{\nu+1}+2\delta^\nu}{\lambda_{\nu+1}+\lambda_\nu}+\frac{3\left(E^{\nu+1}-E^\nu\right)}{\left(\lambda_{\nu+1}+\lambda_\nu\right)^2},\\
c_\nu &= \delta^\nu,\\
d_\nu &= E^\nu.
\end{align}
The derivatives of the energy, $\delta_\nu$, provide important information about the path. 
In some cases, intermediate minima on the MEP can be identified which could not be
seen when only values of the energy are used in the interpolation. 
When an intermediate minimum is found, it is advisable to split the path
and calculate the two parts of the MEP separately.

The orientation of the magnetic vectors at the saddle point can be obtained from the value of $\lambda$ at the maximum of the energy curve.



\section{Energy surface for the test calculations}
\label{app:test_surf}

For the simple test problem calculations, 
the energy of the system is taken to be a sum of Gaussian functions of the geodesic distances
\begin{equation}
\label{eq:gaussian}
E(\vec{m}) = \sum_{j=1}^D a_j\exp\left[-\frac{l_j^2(\vec{m})}{2\sigma_j^2}\right],
\end{equation}
where $l_j(\vec{m})$
is the geodesic distance between the endpoint of vector $\vec{m}$ and a center of the $j$-th Gaussian, $\vec{m_j}$. Parameters defining the center of Gaussians, $\vec{m_j}$, their amplitude, $a_j$, and width, $\sigma_j$ were chosen so as to form multiple minima, maxima and saddle points on the energy surface. 

The values of the parameters of the Gaussian functions, i.e. the amplitude ($a_i$), width ($\sigma_i$) and position of the center 
(in spherical coordinates $\theta_i$ and $\phi_i$) used in calculations shown in Figs. 2 and 3 are listed in the table below:
\begin{center}
\item[]\begin{tabular}{|c|cccc|}
\hline
$i$ & $a_i$ & $\sigma_i$ & $\theta_i$ & $\phi_i$ \\
\hline
1 & -2.500 & 0.200 & 0.600 & 3.700 \\
\hline
2 & -2.000 & 0.150 & 0.650 & 6.300 \\
\hline
3 & 4.000 & 0.300 & 0.150 & 5.000 \\
\hline
4 & 0.656 & 0.080 & 1.523 & 4.280 \\
\hline
5 & 1.609 & 0.341 & 1.662 & 4.899 \\
\hline
6 & 0.696 & 0.303 & 1.753 & 3.963 \\
\hline
7 & 1.753 & 0.488 & 1.404 & 5.777 \\
\hline
8 & 0.429 & 0.276 & 0.132 & 5.655 \\
\hline
9 & 1.659 & 0.862 & 1.831 & 3.974 \\
\hline
10 & 1.222 & 0.568 & 1.291 & 4.539 \\
\hline
11 & 1.302 & 0.402 & 0.020 & 1.968 \\
\hline
12 & 1.685 & 0.743 & 2.494 & 1.141 \\
\hline
13 & 1.287 & 0.800 & 1.438 & 3.046 \\
\hline
14 & 0.249 & 0.282 & 0.780 & 2.869 \\
\hline
15 & 0.304 & 0.443 & 1.609 & 0.611 \\
\hline
16 & 0.656 & 0.430 & 1.361 & 4.123 \\
\hline
17 & 0.057 & 0.601 & 2.260 & 2.854 \\
\hline
18 & 1.337 & 0.236 & 2.245 & 2.135 \\
\hline
19 & 0.075 & 0.813 & 0.561 & 5.211 \\
\hline
20 & 0.652 & 0.776 & 2.232 & 4.565 \\
\hline
21 & 0.392 & 0.453 & 1.741 & 4.825 \\
\hline
22 & 0.014 & 0.565 & 0.860 & 3.719 \\
\hline
23 & 0.111 & 0.484 & 1.648 & 0.490 \\
\hline
24 & 0.884 & 0.160 & 2.867 & 5.354 \\
\hline
%
\end{tabular}
\label{tab:1a}
\end{center}



\section{Iterative optimization algorithm}
\label{app:vpo}

The GNEB force needs to be zeroed by moving down-hill in energy towards an MEP. 
Although many techniques are available for such an optimization problem, standard iterative algorithms where the search direction at each step is defined by the force are not efficient for curved manifolds such as $\mathcal{R}$.
Even though the force defined in Eq.~(\ref{eq:forceGNEB}) is correctly confined to the local tangent space, any finite displacement along a straight line in the direction of the force brings the system out of the $\mathcal{R}$-manifold 
and the magnetic constraints then need to be restored at each iteration. It is preferable to advance the system along a geodesic of the curved 
$\mathcal{R}$-manifold~\cite{abrudan_08, edelman_98}. 
We have used a simple but efficient method based on an equation of motion where the velocity is damped by including only the
component in the direction of the force. The method is referred to here
as velocity projection optimization (VPO). It is analogous to a method 
that has been used in NEB calculations for unconstrained Euclidean configuration space~\cite{NEBleri}. 
Here, it is adapted to magnetic systems by advancing the system along geodesics, thereby automatically fulfilling the magnetic constraints. 
The details are given in the following subsection.  Some special considerations arise in the regions close to the poles. A method for resolving those
issues is described in the second subsection. 


\subsection{Velocity projection optimization on a sphere}
In the VPO method, the coordinates and velocities are updated according to equations of motion. 
In every iteration, only the component of the velocity parallel to the force is kept, 
unless it is pointing in a direction opposite to the force, at which point it is zeroed. 
The advantage of this formulation is that if the system moves in the right direction, it accelerates and approaches faster the region of an optimum. 
If the algorithm overshoots and the system has gone passed the optimum, the velocity is zeroed.

In order to be able to use the VPO method for magnetic systems, the evolution of the system has to satisfy the magnetic constraints. 
While the Landau-Lifshitz equations describe the spin dynamics, they cannot be used in this optimization algorithm since they
do not involve mass and acceleration. 
Instead, the motion of a point mass on the manifold $\mathcal{R}$ is used here. 
This motion is conveniently described in terms of spherical coordinates, $\theta_i$ and $\phi_i$, which automatically account for the curvature of the manifold $\mathcal{R}$. For a system of $N$ magnetic moments, the following set of $2N$ coupled equations are used:
\begin{align}
m\ddot{\theta}_i-m\dot{\phi}_i^2\sin\theta_i\cos\theta_i&=f_i^\theta,\label{eq:pend_1}\\
m\ddot{\phi}_i\sin\theta_i+2m\dot{\phi}_i^2\cos\theta_i&=f_i^\phi,\label{eq:pend_2}
\end{align}
Here, $m$ is the mass, ($\theta_i$,$\phi_i$) are polar and azimuthal angles defining the orientation of $i$-th magnetic moment, 
and ($f_i^\theta$,$f_i^\phi$) are projections of the $i$-th force on the orthogonal unit vectors, ($\hat{e}^\theta_i$,$\hat{e}^\phi_i$), 
in the direction of increasing ($\theta_i$,$\phi_i$). 
These equations are similar to the equations of motion for a spherical pendulum~\cite{landau_76}. 
Eqs.~(\ref{eq:pend_1}), (\ref{eq:pend_2}) are then transformed to coupled first order equations for position and velocity:
\begin{align}
\dot{v}_i^\theta &=v^\phi_i v^\phi_i\cot\theta_i +\frac{1}{m}f_i^\theta,\label{eq:pend_v1}\\
\dot{v}_i^\phi &=-v^\theta_i v^\phi_i\cot\theta_i+\frac{1}{m}f_i^\phi,\label{eq:pend_v2}\\
\dot{\theta}_i &= v^\theta_i,\label{eq:pend_x1}\\
\dot{\phi}_i &= \frac{v^\phi_i}{\sin\theta_i},\label{eq:pend_x2}
\end{align}
These equations are solved iteratively using a numerical scheme. 
We have found that even the simplest Euler integrator works well. 

At each time step, the velocity is modified depending on the sign of its projection on the direction of the force. 
More specifically, if $\bm{\EuScript{F}}$ is a $2N$-dimensional vector of the force with components 
$f_1^\theta,\ldots,f_N^\theta,f_i^\phi,\ldots,f_N^\phi$, and $\bm{\EuScript{V}}=\left(v^\theta_1,\ldots,v^\theta_N,v^\phi_1,\ldots,v^\phi_N\right)$ is a $2N$-dimensional velocity vector, then 
\begin{equation}
\label{eq:vel_mod}
\bm{\EuScript{V}}^{rev}=
\begin{cases}
\left(\bm{\EuScript{V}}\cdot\bm{\EuScript{F}}\right)\bm{\EuScript{F}}/\left|\bm{\EuScript{F}}\right|^2, &\text{if $\left(\bm{\EuScript{V}}\cdot\bm{\EuScript{F}}\right)>0$,}\\
0, &\text{if $\left(\bm{\EuScript{V}}\cdot\bm{\EuScript{F}}\right)\le 0$.}
\end{cases}
\end{equation}
Here, superscript 'rev' stands for 'revised'. The revised velocity $\bm{\EuScript{V}}^{rev}$ is used in the following iteration. 
If the force is consistently pointing in a similar direction, the system accelerates in that direction. 
This is equivalent to increasing the timestep in a steepest descent minimization. 
Then, when the system overshoots the optimum, $(\bm{\EuScript{V}}\cdot\bm{\EuScript{F}})\le 0$, the velocity is zeroed.

This procedure 
is applied simultaneously to all images in the elastic band. $f_i^\theta$ and $f_i^\phi$ can be derived from Cartesian components of the force using the following relations~\cite{korn_00:2}:
\begin{align}
f_i^\theta &= f_i^x\cos\theta_i\cos\phi_i+f_i^y\cos\theta_i\sin\phi_i-f_i^z\sin\theta_i, \label{eq:cart_spher1}\\
f_i^\phi &= -f_i^x\sin\phi_i+f_i^y\cos\phi_i.\label{eq:cart_spher2}
\end{align}
When the VPO method is used to find an energy minimum, the force is simply the negative of the energy gradient with respect to spherical coordinates~\cite{korn_00:2}.
The VPO optimization typically converges to the nearest local energy minimum. 
The global energy minimum can in principle be found by sampling enough initial guesses and repeating the procedure.


\subsection{Treatment of the poles}
During the force minimization, magnetic vectors may approach the vicinity of the poles where the azimuthal angle $\phi$, 
becomes irrelevant. In this case the VPO formulated in terms of spherical coordinates will break down. 
Therefore, some other coordinate system needs to be chosen in the region of the poles.
One option is to use the $x$ and $y$ components of magnetic moments. 
More specifically, if $\theta_i<\epsilon$ or $\pi-\theta_i<\epsilon$, where $\epsilon$ is some small positive number, then the 
coordinates $m_i^x$ and $m_i^y$ should be used instead of $\theta_i$ and $\phi_i$ in the VPO procedure:
\begin{align*}
m_i^x&=\sin\theta_i\cos\phi_i,\\
m_i^y&=\sin\theta_i\sin\phi_i.
\end{align*}
Furthermore, the velocity also needs to be expressed in terms of Cartesian coordinates~\cite{korn_00:2}:
\begin{align}
v_i^x &= v_i^\theta\cos\theta_i\cos\phi_i-v_i^\phi\sin\phi_i,\\
v_i^y &= v_i^\theta\cos\theta_i\sin\phi_i+v_i^\phi\cos\phi_i.
\end{align}
The evolution of a point in the $\epsilon$-neighborhood of the pole is then governed by the usual Newton's equations of motion. 
The  $f_i^\theta$, $f_i^\phi$ and $v_i^\theta$, $v_i^\phi$ components 
of the $2N$-dimensional force and velocity vectors, $\bm{\EuScript{F}}$ and $\bm{\EuScript{V}}$, 
should be substituted by $f_i^x$, $f_i^y$ and $v_i^x$, $v_i^y$ in the velocity projection of Eq.~(\ref{eq:vel_mod}). 
When the system leaves the $\epsilon$-neighborhood of the pole, 
inverse transformation to spherical coordinates needs to be performed~\cite{korn_00:2}.

\begin{figure}[h!]
\centering
\includegraphics[width=0.3\columnwidth]{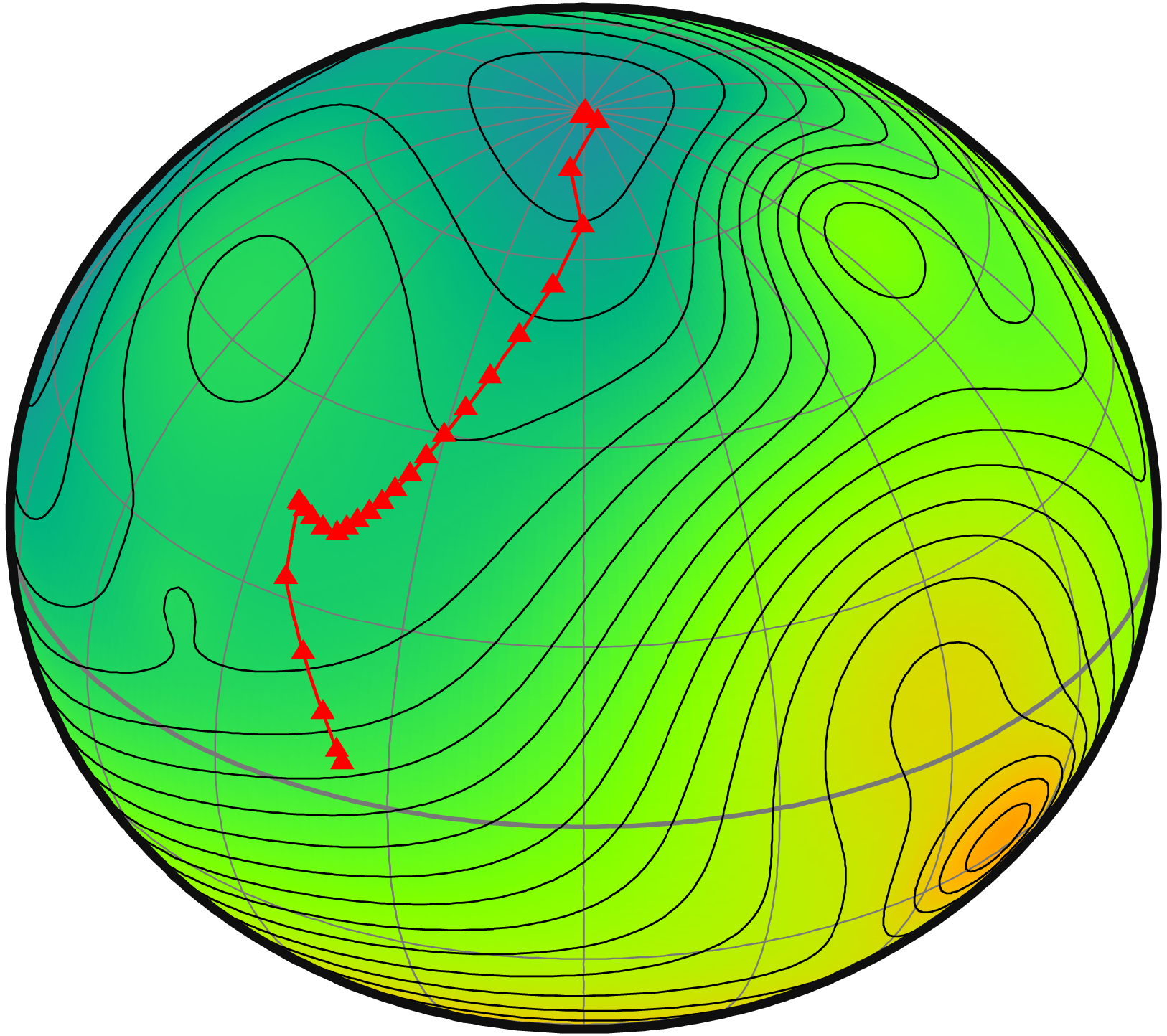}
\caption{
Minimization of the energy of a test system using the VPO algorithm as described in Appendix~\ref{app:vpo}. 
The trajectory has a cusp where the inner product of velocity and force becomes negative and the velocity is zeroed. 
The minimization path changes direction at that point and the system advances slowly from there at first 
but then accelerates as the force keeps pointing in a similar direction. 
The trajectory converges at the minimum which is located at the pole without a problem. 
}
\label{fig6}
\end{figure}

A typical trajectory of the representative point during the VPO calculations is shown in Fig.~\ref{fig6}. 
This trajectory has a cusp where the inner product of velocity and force becomes negative and the velocity is zeroed. 
The minimization path changes direction at that point and subsequently the system advances slowly from at first 
but then accelerates as the force keeps pointing in a similar direction.
The trajectory converges at the minimum which is located at the pole without a problem. 

\section{Verification of a saddle point}
\label{app:verif_sp}

After a maximum along the MEP has been found, either by interpolation of the path between adjacent images or by using the CI-GNEB procedure, 
it is important to check whether this maximum indeed corresponds to a first order SP on the energy surface.
This analysis is in any case needed to evaluate the transition rate using either HTST or Kramers methods.
A first order SP is an extremum on the energy surface which is a maximum with respect to one and only one eigenmode 
while it is a minimum with respect to the other modes. In some cases, when a highly symmetrical initial path is used, 
the NEB can result in a path going through a higher order SP, i.e. a stationary points on the energy surface which is a maximum
with respect to two or more modes. 
Energy surfaces can also be constructed where a ridge comes directly in continuation of an energy valley and the NEB method can then
converge on a path lying partly along the ridge~\cite{Sheppard_11}.
For simple systems with only a few degrees of freedom, as the ones considered in the test problems, 
the energy surface and the MEP can be visualized, so it is quite easy to check the results of the calculations. 
For complex magnetic systems with multiple 
magnetic vectors, a reliable strategy for verification of a SP is to calculate the $2N\times2N$ matrix of the second derivatives of the energy, the Hessian matrix, and calulate its eigenvalues. Any convenient set of coordinates can be used when computing the second derivatives, including spherical coordinates, stereographic coordinates, etc. 
If one and only one eigenvalue of the Hessian matrix is negative, 
then the maximum along the converged path is indeed a first order SP. 
If two eigenvalues are negative, then it is a second order saddle point, etc. 
If the NEB converges on a path going through a second order saddle point, 
then the highest energy image can be displaced in a direction along the negative mode that is orthogonal to the path tangent.
Another option is to apply the minimum mode following method 
starting from the highest energy image~\cite{henkelman_99,olsen_04}.
Convergence to a path going through a higher order saddle point can also be the result of accidental symmetry in the initial path.
It is in general good to add small random noise to the initial path to break such symmetries.



\end{document}